\documentclass[12pt,a4paper]{article}
\usepackage{fullpage}
\usepackage[round]{natbib}

\usepackage[utf8]{inputenc}

\usepackage{amsmath,amsfonts,amssymb}
\usepackage{bm}
\usepackage{mathtools} %
\usepackage{xfrac}
\usepackage{siunitx}

\usepackage{hyperref}
\usepackage[usenames,dvipsnames,svgnames,table]{xcolor}
\usepackage{graphicx}
\usepackage{lmodern}

\usepackage[ruled]{algorithm2e} %

\usepackage{stmaryrd} %
\SetSymbolFont{stmry}{bold}{U}{stmry}{m}{n} %

\pdfoutput=1
\usepackage{hyperref}
\hypersetup{
    colorlinks=true,
    linkcolor=black,
    citecolor=black,
    filecolor=black,
    urlcolor=black,
}
\usepackage[noabbrev]{cleveref}

\newcommand{\T}[1]{{\boldsymbol{{#1}}}}
\newcommand{\Op}[1]{{\mathcal{{#1}}}}
\newcommand{\M}[1]{{\mathbf{{#1}}}}
\newcommand{\V}[1]{{\mathbf{{#1}}}}

\newcommand{\pdiff}[2]{\frac{\partial#1}{\partial#2}}

\newcommand{\dif}{\,\mathrm{d}}

\newcommand{\jump}[1]{\left\llbracket\,#1\,\right\rrbracket}
\newcommand{\avg}[1]{\left\{\!\left\{ #1 \right\}\!\right\}}

\newcommand{\Hdiv}{H\textsuperscript{div}}

\newcommand{\vel}{\T{u}}
\newcommand{\wvel}{\T{w}}
\newcommand{\vtest}{\T{v}}
\newcommand{\cfunc}{\alpha}
\newcommand{\ctest}{\beta}
\newcommand{\pos}{\T{x}}
\newcommand{\grav}{\T{g}}

\newcommand{\element}{K} %

\newcommand{\tess}{\mathcal{T}}
\newcommand{\skel}{\mathcal{S}}

\newcommand{\Co}{\mathrm{Co}}

\newcommand{\normal}{\T{n}}

\usepackage{listings}
\usepackage{caption}
\crefname{lstlisting}{listing}{listings}
\Crefname{lstlisting}{Listing}{Listings}
\lstdefinelanguage{yaml}{
  basicstyle=\footnotesize\color{black}\ttfamily,
}

\begin{document}

\iffalse
\begin{frontmatter}

  \journal{Computer Physics Communications}

  \title{COPY FROM BELOW}
  
  \author{Tormod Landet\corref{author}}
  \author{Mikael Mortensen}
  
  \cortext[author]{Corresponding author.\\\textit{E-mail address:} tormodla@math.uio.no}
  \address{Department of Mathematics, Division of Mechanics, University of Oslo, Blindern 0316 Oslo, Norway}
    
  \maketitle

  \begin{abstract}
    COPY FROM BELOW
  \end{abstract}


\end{frontmatter}

\noindent\textbf{Program summary}\\
\begin{small}
  \noindent
  {\em Program Title:} Ocellaris\\
  {\em Licensing provisions:} Apache License, Version 2.0\\
  {\em Programming language:} Python 3 and C++14\\
  {\em Supplementary material:} Source code, input files and 
   Docker/Singularity container recipes to reproduce the presented results are available for download.\\
  {\em Nature of problem:} The simulation of free-surface flows at high Reynolds numbers, such as ocean waves, are important for the design of marine and coastal structures. The numerical approximation of two-phase flows with large and sharp transitions in the density field is challenging when using higher-order methods due to convective instabilities creating spurious Gibbs oscillations in the solution. Accurate free-surface models require exact incompressibility, which is non-trivial to combine with convective stability in a higher-order method.\\
  {\em Solution method:} We use an exactly divergence-free interior-penalty Discontinuous Galerkin (DG) finite element discretisation with a special slope-limiting strategy for the velocity field to obtain stable and high-order solutions while retaining exact mass conservation. This is coupled with a volume-of-fluid (VOF) method for density-field advection and free-surface capturing. The Navier--Stokes solver is parallelised by use of an incremental algebraic pressure-splitting scheme that produces an exactly divergence-free velocity field independently of the number of pressure-correction iterations.\\
  {\em Additional comments including Restrictions and Unusual features:} The numerical method is designed to be as simple as possible to show that higher-order exactly incompressible DG methods are feasible and can be used to simulate realistic high Reynolds number free-surface flows with large density differences without requiring complicated numerical schemes or parameter tuning. As a result, particularly the free-surface capturing method could benefit greatly from relatively simple improvements such as using a nested sub-mesh for the VOF colour function, instead of throwing away the higher-order properties when advecting the density field. Ocellaris is written to be highly modular; features such as the VOF model, boundary conditions or the Navier--Stokes solver itself can be replaced by the user in the simulation input file without changing the Ocellaris source code itself. The input file format is fully described in Ocellaris' user guide.
\end{small}

\else

\title{Ocellaris: a discontinuous Galerkin finite element solver for two-phase flows with high density differences}
\author{
  Tormod Landet%
  , %
  Mikael Mortensen
  \\%
  \vspace{6pt}%
  {\em{Department of Mathematics, University of Oslo}}%
  \vspace{-2mm}\\%
  {\em{Moltke Moes vei 35, 0851 Oslo, Norway}}
}
\maketitle

\begin{abstract}
  In free-surface flows, such as breaking ocean waves, the momentum field will have a discontinuity at the interface between the two immiscible fluids, air and water, but still be smooth in most of the domain. Using a higher-order numerical method is more efficient than increasing the number of low-order computational cells in areas where the solution is smooth, but higher-order approximations cause convective instabilities at discontinuities. In Ocellaris we use slope limiting of discontinuous Galerkin solutions to stabilise finite element simulations of flows with large density jumps, which would otherwise blow up due to Gibbs oscillations resulting from approximating a factor 1000 sharp jump (air to water) by higher-order shape functions.

  We have previously shown a slope-limiting procedure for velocity fields that is able to stabilise 2D free-surface simulations running on a single CPU. In this paper our solver is extended to 3D and coupled to an algebraic pressure-correction scheme that retains the exact incompressibility of the direct solution used in the 2D simulations. We have tested the method on a common 3D dam-breaking test case and compared the free-surface evolution and impact pressures to experimental results. We also show how a forcing-zone approach can be used to simulate a surface-piercing vertical cylinder in an infinite wave field. In both cases the free-surface elevation and the forces and pressures compare well with published experiments. The Ocellaris solver is available as an open-source and well-documented program along with the input files needed to replicate the included results (\href{https://www.ocellaris.org/}{www.ocellaris.org}).
\end{abstract}
\fi

\section{Introduction}
\label{sec:intro}

Increasing the polynomial approximation order is more efficient than increasing the number of computational cells when solving partial differential equations (PDEs) where the solution is smooth \citep{babuska_hp_error_1981}. Performing more work locally in higher-order cells is also advantageous for today's parallel computers \citep{Kubatko09,Kirby12,Huerta13}. However, for equations with convective operators, a jump in the solution or the coefficients will cause nonlinear convective instabilities---spurious Gibbs oscillations at discontinuities---that will eventually destroy the solution if the ratio between the high and low sides of the jump is large. Linear numerical schemes, such as the discontinuous Galerkin finite element method (DG FEM), produces linear algebraic equations from discretisation of linear PDEs. Such linear schemes cannot guarantee convective stability if they are not monotonic, and a linear monotonic scheme must be first-order and is hence highly dissipative \citep{krivodonova_shock_2004,harten_high_resolution_schemes_1983}. One way to get around the problem of having to chose between high-order approximations and convective stability is to make the scheme nonlinear by the inclusion of a nonlinear cell-wise projection operator, a slope limiter.

We have previously shown that component-wise slope limiters can be used on the convected velocity field to stabilise two-phase flow simulations with large density jumps \citep{landet_slopelim_2019}. Most existing methods used for the simulation of air/water two-phase flows employ low-order finite volume methods for the discretisation of the governing PDEs. In these methods the solution is piecewise constant and the convective instabilities can be dealt with by using a flux limiter, a non-linear facet-local parameter that blends the upwind and downwind fluxes to obtain stability without excessive diffusion \citep{hirt_volume_1981,ubbink_1997,weller_FOAM_1998,popinet_gerris_2003,kleefsman_vof_2005}.

Ocellaris solves the variable-density Navier--Stokes equations for two-phase flow (\cref{sec:goveq}). The numerical method is based on a higher-order interior-penalty DG FEM (\cref{sec:dgmethod}) with a component-wise velocity slope limiter for the convected velocity field (\cref{sec:slopelim}) and an algebraic pressure-correction method for the pressure--velocity coupling (\cref{sec:algo}). A regular wave model based on a stream-function formulation is used for initial and boundary values for water-wave simulations. The details of this model are presented in \cref{sec:waves}, including how the stream-function approach can be used to specify one velocity field, valid in both the air and water phases, which is divergence free, satisfies all boundary conditions, and conforms to the free surface. \Cref{sec:waves} also describes how a forcing-zone approach is used for damping free-surface disturbances near the boundaries in order to avoid unwanted reflections. The implementation of Ocellaris is briefly described in \cref{sec:implementation}, and the main steps required to set up an Ocellaris simulation are described in \cref{sec:inpfile}.

The \texttt{BlendedAlgebraicVOF} multiphase model used in this paper computes a piecewise constant density distribution based on an algebraic VOF method \citep{muzaferija_hric_1998}. Other multiphase models are available, and it is also possible to define a custom model in the Ocellaris simulation input file (YAML format, see \citet{landet_ocellaris_user_guide_2019} for details). Using such a very simple model is not optimal, but the velocity slope limiter will lower the effective order of the obtained velocities at the density jump, so perhaps not much is lost in terms of obtainable accuracy from using a relatively simple free-surface capturing method. An interface-capturing method that can take full advantage of the high order of the convecting velocity field is an interesting research topic, and would most likely require fewer cells in the free-surface region than what is used here. Still, the results show good agreement with lab experiments, and away from the free surface the overall method retains the high-order approximation properties of the Navier--Stokes discretisation since the true density field is constant here.

\Cref{sec:dambreak} shows the performance of the Ocellaris solver on a 3D dam-breaking test case. The test is meant to simulate a ``green-water'' event, a large wave breaking over a ship deck and impacting the cargo, in this case a single container outfitted with pressure gauges. The results show that both the free-surface evolution and the impact pressures compare well with published experiments. The second test case, presented in \cref{sec:cylinder}, is a vertical cylinder exposed to steep regular waves. This is meant to simulate wave loads on an offshore windmill or another type of slender surface-piercing marine structure. This test shows good agreement in regard to the total force on the cylinder when compared to laboratory experiments. The test also shows that the forcing-zone approach presented in \cref{sec:waves} works well in a DG FEM setting to dampen free-surface disturbances near the inlet and outlet of the numerical wave tank. Finally, the discussion in \cref{sec:discussion} concludes that Ocellaris' high-order multiphase flow solver can successfully simulate complex 3D air/water free-surface flows at high Reynolds numbers.

\section{Mathematical model of free-surface flow}
\label{sec:goveq}

Ocellaris solves the variable-density Navier--Stokes equations \labelcref{eq:nsmom,eq:nssol,eq:nsdens} with piecewise constant density and viscosity. Standard notation is used for the unknown functions; $\vel$ is the velocity, $p$ is the pressure, and $\rho$ is the fluid density. The coefficients are the dynamic viscosity $\mu$ and the acceleration of gravity $g$,
\begin{align}
  \rho \left( \pdiff{\vel}{t} + (\vel\cdot\nabla) \vel \right) &= \nabla\cdot\mu\left(\nabla \vel + (\nabla\vel)^T\right) - \nabla p + \rho \grav, \label{eq:nsmom}\\
  \nabla\cdot \vel &= 0, \label{eq:nssol}\\
  \pdiff{\rho}{t} + \vel\cdot\nabla \rho &= 0. \label{eq:nsdens}
\end{align}

Ocellaris is currently designed to study air/water free-surface flows at high Reynolds numbers, and does not include a turbulence closure model or the effect of surface tension. The error made by neglecting these effects is small for the included benchmark tests \citep{kleefsman_vof_2005,paulsen_forcing_2014}. The program design is made to be easily extendable and there should be no fundamental problem with including such effects at a later time. For the first public release, the sole focus of the Ocellaris project has been to show that using higher-order DG methods for free-surface flows is feasible and can be made stable in regard to convective instabilities without compromising mass conservation.

A VOF colour function approach is used for density transport and free-surface capturing \citep{hirt_volume_1981},
\begin{align}
 \rho = \cfunc\,\rho_\text{water} + (1 - \cfunc)\,\rho_\text{air},
\end{align}
where $\cfunc\in[0,1]$ is the colour/indicator function which is used as the unknown instead of the fluid density. The kinematic viscosity, $\nu=\sfrac{\mu}{\rho}$ is computed similarly,
\begin{align}
  \nu = \cfunc\,\nu_\text{water} + (1 - \cfunc)\,\nu_\text{air}.
\end{align}

\section{Discontinuous Galerkin discretisation}
\label{sec:dgmethod}

The numerical method is an extension of \citet{landet_slopelim_2019} which builds on \citet{cockburn_locally_2005} and uses the symmetric interior-penalty (SIP) method for the viscous term \citep{arnold_interior_1982}. The domain is discretised as an irregular mesh comprised of tetrahedral cells. Let $\tess$ be the set of all cells and $\skel$ the set of all facets in the mesh. Polynomial function spaces of degree $k$ on each cell $\element$ are denoted $P_k(\element)$. These have no continuity at cell boundaries and no inherent boundary conditions.

We use calligraphic typeface to denote operators and sets, bold italic for vectors functions and italic for scalar functions. For nabla the conventions $(\nabla\vel)_{ij}=\partial_j\vel_i$ and $(\nabla\cdot\T{\sigma})_i=\partial_j\sigma_{ij}$ are used. For time derivatives $\vel=\vel^{n+1}$ is the unknown trial function and $\vel^{n}$ and $\vel^{n-1}$ are the known values at the previous two time steps.
A second-order backwards-differencing formulation, BDF2, is used for time integration. The parameters are $\{\gamma_1, \gamma_2, \gamma_3\} = \{\sfrac{3}{2}, -2, \sfrac{1}{2}\}$, see the first integral in \cref{eq:wf_ns_coupled1}. The convecting velocity $\wvel$ is considered known---which linearises the momentum equation---and $\wvel$ is \Hdiv-conforming such that the flux is continuous, $\jump{\T{w}}_\normal=0$, see \cref{sec:algo} for details.

The governing equations, \labelcref{eq:nsmom,eq:nssol,eq:nsdens}, are cast into the following form: find $\vel\in [P_2(\element)]^3$, $p\in P_{1}(\element)$, and $\cfunc\in P_0(\element)$ such that
\begin{alignat}{4}
\Op{A}(\vel, \vtest; \wvel) + \Op{B}(p, \vtest) &= \Op{D}(\vtest) &&\forall\ \vtest\ &&\in [P_2(\element)]^3, \label{eq:op_nsmom}\\
\Op{C}(\vel, q) &= \Op{E}(q)                                      &&\forall\ q       &&\in P_{1}(\element), \label{eq:op_nssol} \\
\Op{F}(\cfunc, \ctest; \wvel) &= \Op{G}(\ctest)     \quad         &&\forall\ \ctest  &&\in P_{0}(\element), \label{eq:op_nsdens}
\end{alignat}
in the tessellated domain $\tess$ subject to
\begin{alignat}{2}
\vel &= \vel_D && \quad\text{on Dirichlet boundary facets, } \skel_D\subset\skel \\
\pdiff{\vel}{\normal} &= \T{a} && \quad\text{on Neumann boundary facets, } \skel_N\subset\skel
\end{alignat}

The discontinuous Galerkin (DG) method works by breaking integrals over the whole domain into a sum of integrals over each mesh cell $\element\in\tess$, and defining fluxes of the unknown functions between these cells. The average and jump operators across an internal facet between two cells $\element^+$ and $\element^-$ are defined as 
\begin{align}
\avg{u} &= \frac{1}{2}(u^+ + u^-), \\
\jump{u} &= u^+ - u^-, \\
\jump{\T{u}}_\normal &= \T{u}^+\cdot\normal^+ + \T{u}^-\cdot\normal^-.
\end{align}
where $u^+$ is the value of $u$ along the internal facet when computed using the shape functions and degrees of freedom related to the $\element^+$ cell and vice versa for $u^-$.
For the exterior facets, which only have one connected cell due to being located on the domain boundary, let the connected cell be denoted $\element^+$ such that $\normal^+\!\cdot\jump{\vel} = \normal^+\!\cdot\vel^+ = \normal\cdot\vel$. Take $\avg{u}=u$ and otherwise let all $\element^-$ values related to the non existing element outside the domain be zero.

\subsection{Momentum equation}

The momentum \cref{eq:nsmom} is discretised using the SIP method for the elliptic term \citep{arnold_interior_1982} and otherwise using the fluxes from \citet{cockburn_locally_2005}. The application of boundary conditions and the choice of the penalty parameter $\kappa_\mu$ is described in detail in \citet{landet_slopelim_2019}. The flux of pressure is $\hat{p}=\avg{p}$ and the convective flux $\hat\vel^\wvel$ is a pure upwind flux. After multiplication with $\vtest$ and integration over $\tess$, followed by integration by parts and application of the SIP method to the viscosity, $\Op{A}$ can be found as the bilinear part containing $\vel=\vel^{n+1}$, $\Op{B}$ as the bilinear part containing $p$, and $\Op{D}$ as the linear part of

{\allowdisplaybreaks
\begin{flalign}
&\int_\tess\frac{\rho}{\Delta t}(\gamma_1\vel + \gamma_2\vel^{n} + \gamma_3\vel^{n-1})\vtest\dif\pos
\label{eq:wf_ns_coupled1}
\\ \notag
&\qquad-\ \int_\tess \vel \cdot\nabla\cdot (\rho\vtest\otimes\wvel)\dif\pos
 \ +\ \int_\skel \wvel\cdot\normal^+\,\hat\vel^\wvel\cdot\jump{\rho\vtest}\dif s
\\ \notag
&\qquad+\ \int_\tess \mu\left(\nabla\vel + (\nabla\vel)^T\right):\nabla\vtest\dif\pos
\ +\ \int_{\skel_I}\kappa_\mu\jump{\vel}\cdot\jump{\vtest}\dif s
\\ \notag
&\qquad-\ \int_\skel (\avg{\mu\left(\nabla\vel + (\nabla\vel)^T\right)}\cdot\normal^+)\cdot\jump{\vtest}\dif s
\\ \notag
&\qquad-\ \int_{\skel_I} (\avg{\mu\left(\nabla\vtest + (\nabla\vtest)^T\right)}\cdot\normal^+)\cdot\jump{\vel}\dif s
\\ \notag
&\qquad-\ \int_\tess p\, \nabla\cdot\vtest\dif\pos
 \ +\ \int_\skel\hat{p}\,\normal^+\cdot\jump{\vtest}\dif s
\ =\ \int_\tess\rho\,\grav\dif\pos.
\end{flalign}}

On Dirichlet boundaries $\hat\vel^\wvel$ is the upwind value. Depending on flow direction this is either $\vel$ or $\vel_D$.  On Neumann boundaries let $\hat\vel^\wvel=\vel$. The boundary conditions (BCs) can be set separately for each velocity component, allowing slip, $\vel\cdot\normal=0$, or non-slip, $\vel=0$, BCs. On all boundaries take $\hat{p}=p$.

\subsection{Continuity equation}

The equation used to ensure a divergence-free velocity field \eqref{eq:nssol} is multiplied by $q$ and integrated over $\tess$. The flux is $\hat{\vel}^p=\avg{\vel}$. After integration by parts $\Op{C}(\vel,q)$ and $\Op{E}(q)$ can be found from
\begin{align}
&\int_\skel \hat{\vel}^p\cdot\normal^+\jump{q} \dif s
 \ -\ \int_\tess \vel\cdot\nabla q\dif\pos
 \ =\ 0.
\label{eq:wf_ns_coupled2}
\end{align}

The non-zero $\Op{E}(q)$ results from using the boundary conditions in the flux, $\hat\vel^p=\vel_D$ on Dirichlet boundaries. On Neumann boundaries  the unknown function is used directly, $\hat\vel^p=\vel$, but when assembling the contribution from each velocity component separately it is beneficial to set $\hat\vel^p=0$ on free-slip surfaces also for components with Neumann BCs. Otherwise, it is possible to get $\vel\cdot\normal \neq 0$ due to small errors in the generated meshes when the boundaries are not perfectly approximated by simplices.

\subsection{Density transport}

The HRIC method \citep{muzaferija_hric_1998} is used to define the colour function flux $\hat\cfunc$. The HRIC flux limiter is stable and avoids the excessive interface diffusion of a standard upwind flux. It would be ideal to construct a better interface capturing method that can take advantage of the fact that the velocity is in $[P_2(\element)]^3$, but in this first public release of Ocellaris a standard algebraic VOF method is used to validate the stability and applicability of the overall free-surface flow solver.

To construct the DG operators for the density transport \cref{eq:op_nsdens}, the strong form in \cref{eq:nsdens} is multiplied by the test function $\ctest$, and the result is integrated over the domain. After integration by parts and discarding derivatives of the piecewise constant functions, $\Op{F}$ can be found as the bilinear part and $\Op{G}$ as the linear part of
\begin{align}
&\int_\tess\frac{1}{\Delta t}(\gamma_1 \cfunc^{n+1} + \gamma_2 \cfunc^{n} + \gamma_3 \cfunc^{n-1})\ctest\dif\pos
\label{eq:colour_weak}
+ \int_\skel\hat \cfunc^{n+1}\,\wvel\cdot\normal^+\jump{\cfunc}\dif s = 0,
\end{align}

\subsection{Velocity slope limiter}
\label{sec:slopelim}

Using functions spaces that are higher order than piecewise constants to approximate the velocity will cause Gibbs oscillation instabilities when the density field has jumps, such as the order \num{1000} jump in density in free-surface simulations of air and water. Slope-limiting techniques can be used to remove such instabilities. The velocity slope-limiting approach taken here is the component-wise hierarchical Taylor-polynomial based slope limiter described in \citet{landet_slopelim_2019}, where each velocity component of the convect\kern-0.05em\emph{ed} velocity, $\vel$, is slope limited by a vertex-based scalar slope limiter \citep{kuzmin_vertex-based_2010} while the convect\emph{ing} velocity, $\wvel$, is left unchanged. This approach requires no tuning and the time spent in the velocity slope limiter is only around \SI{0.2}{\percent} of the total running time.

\begin{figure}[htb]
  \centering
  \includegraphics[width=0.45\textwidth]{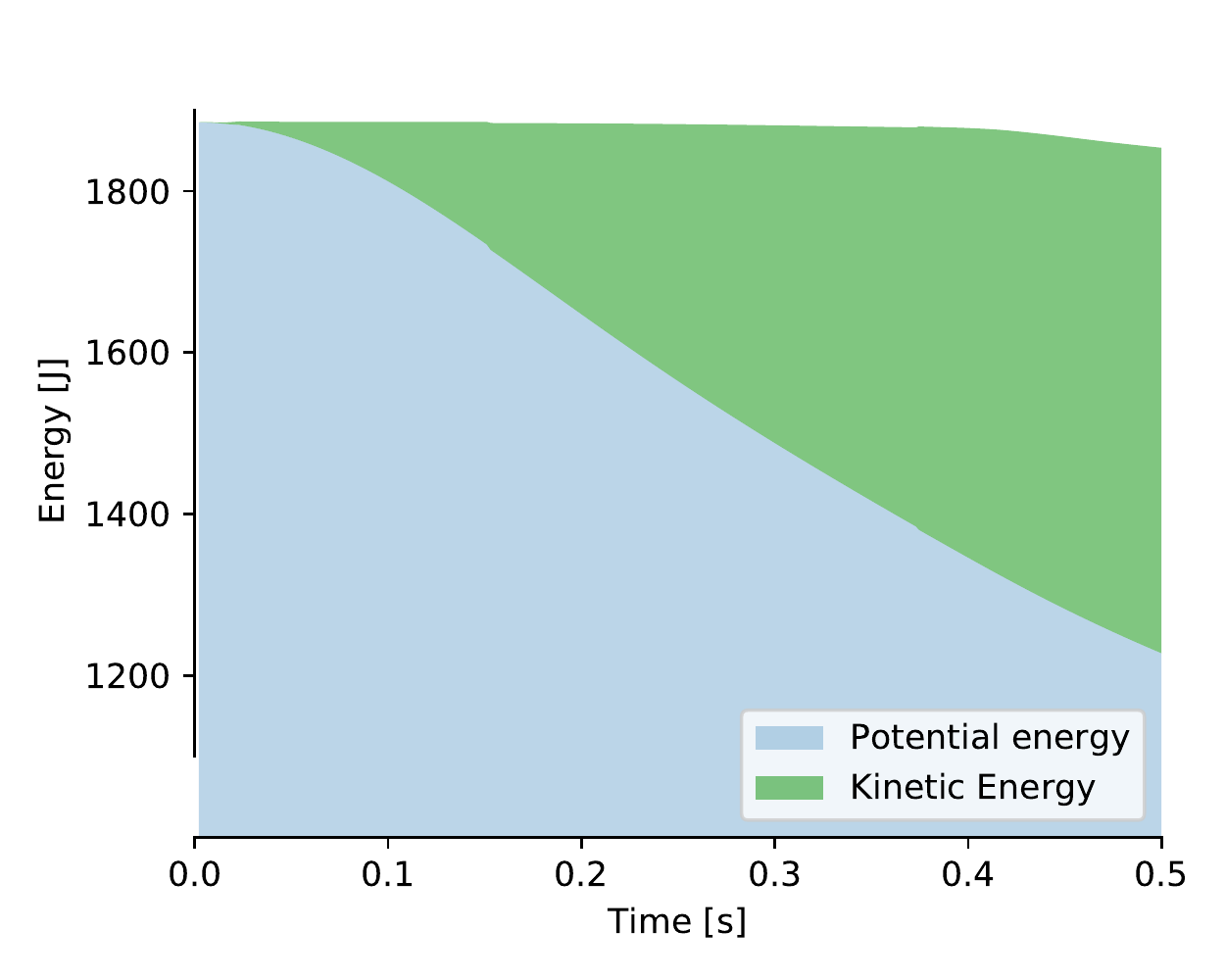} \hspace{0.03\textwidth}
  \includegraphics[width=0.45\textwidth]{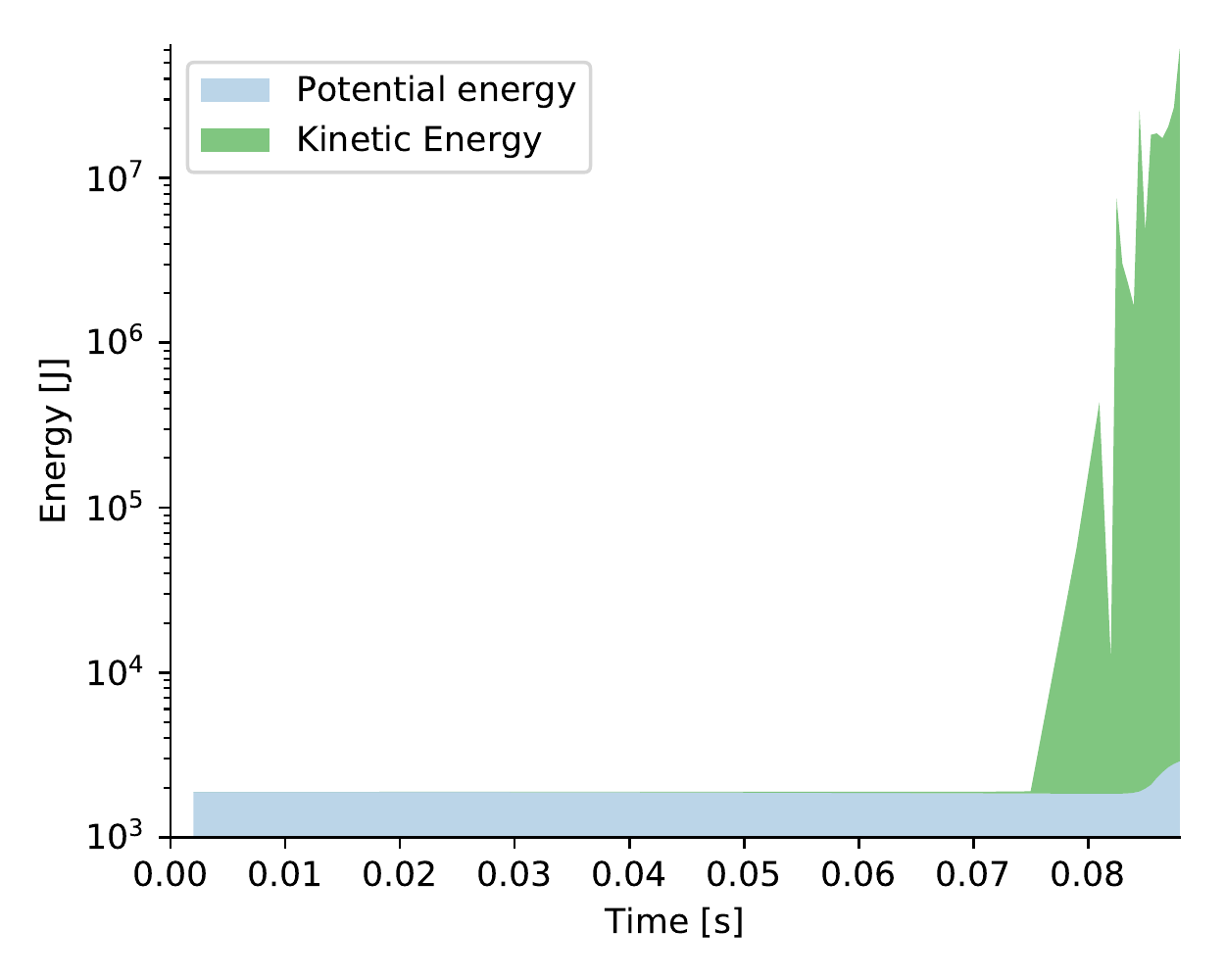}
  \caption{The evolution of the kinetic, potential and total energy. 3D dam breaking, fine mesh, with (left) and without (right) velocity slope limiting. Notice the difference in time scales between the figures.}
  \label{fig:greenwater_energy}
\end{figure}

The effect of using a slope limiter is dramatic. \Cref{fig:greenwater_energy} shows the evolution of the total, potential and kinetic energy in the 3D dam-breaking test case presented more thoroughly in \cref{sec:dambreak}. Water rushes out of the broken dam towards a small box-shaped object which is impacted after about \SI{0.4}{\second}. As can be seen, the slope-limited simulation preserves total energy well, trading potential for kinetic energy up to the time of impact. After the impact the conservation of energy is not perfect,%
but the kinetic energy is always controlled and does not blow up. The second plot shows that the kinetic energy quickly blows up when using the same simulation setup with the velocity slope limiter deactivated. The simulation is automatically stopped after less than \num{50} time steps when the Courant number passes \num{1000}.

\section{Solution algorithm}
\label{sec:algo}

The time-stepping procedure is shown in \cref{algo:timestep}. To decouple the density field solver from the Navier--Stokes solver, and to linearise the momentum equation, a second-order extrapolation $\wvel^*$ is used to estimate the convective velocity $\wvel$,
\begin{equation}
\wvel^* = \wvel^{*,\,n+1} = 2\wvel^{n} - \wvel^{n-1},
\label{eq:explicit_w}
\end{equation}
which means $\rho^{n+1}$ can be computed before the velocity at time $t=(n+1)\Delta t$.

\begin{algorithm}[htb]
  \SetAlgoLined \SetAlgoNoEnd
  \While{$t^{n+1} < t_\mathrm{max}$}{
    Solve for $\rho^{n+1}$ using $\wvel^*$ (estimate of $\wvel^{n+1}$)\;
    \While{not converged}{
      Solve momentum equation for $\vel^*$\;
      Solve pressure correction equation for $p^{n+1}$\;
      Update the velocity $\vel^{**}$ using $\vel^*$ and $p^{n+1}$\;
    }
    Compute $\vel^{n+1}$ by projecting $\vel^{**}$ into BDM\;
    Copy $\vel^{n+1}$ into $\wvel^{n+1}$\;
    Slope limit $\vel^{n+1}$ \;
    Increment $n$\;
  }
  \caption{The Ocellaris IPCS-A time stepping procedure.}
  \label{algo:timestep}
\end{algorithm}

The discretised matrix version of \cref{eq:op_nsmom,eq:op_nssol},
\begin{align}
  \begin{bmatrix}
    \M{A} & \M{B} \\
    \M{C} & 0
  \end{bmatrix}
  \begin{bmatrix}
    \V{u} \\ \V{p}
  \end{bmatrix}
  =
  \begin{bmatrix}
    \V{d} \\ \V{e}
  \end{bmatrix},
  \label{eq:NS_mat_op}
\end{align}
is a saddle-point problem, which is solved by the incremental pressure-correction scheme on algebraic form (IPCS-A). This is an incremental version of the classic Chorin-Temam methods \citep{Chorin_1968,Temam_1969}. $\M{A}$, $\M{B}$ and $\M{C}$ are sparse matrix versions of the $\Op{A}$, $\Op{B}$ and $\Op{C}$ operators, discretised using the discontinuous Galerkin method described in \cref{sec:dgmethod}. The unknowns $\vel$ and $p$ are now vectors of degrees of freedom, $\V{u}$ and $\V{p}$.

The first IPCS-A step is momentum prediction, performed by using an approximate pressure field $\V{p}^*$ inserted into the first row of \cref{eq:NS_mat_op},
\begin{align}
  \M{A}\V{u}^* = \V{d} - \M{B}\V{p^*}.
  \label{eq:IPCSA_mom_with_guess}
\end{align}

A splitting error is now introduced. First let $\M{A}=\M{M}+\M{R}$ where $\M{M}$ is a scaled mass matrix resulting from assembly of the time derivative and $\M{R}$ contains the convective and diffusive operators. Then make the assumption that $\M{R}(\V{u} - \V{u}^*)\approx0$ and use this when subtracting \cref{eq:IPCSA_mom_with_guess} from the first row of \cref{eq:NS_mat_op},
\begin{align}
  \M{M}(\V{u} - \V{u}^*) = - \M{B}(\V{p} - \V{p}^*).
  \label{eq:IPCSA_spliterr}
\end{align}
The matrix $\M{M}$ is block diagonal, and is hence cheap to invert. Use this property and the divergence-free criterion, $\M{C}\V{u} = \V{e}$, to remove the unknown $\V{u}$ from \cref{eq:IPCSA_spliterr}, and reorganise this into an equation for $\V{p}$,
\begin{align}
  \M{C}\M{M}^{-1}\M{B}\V{p} = \M{C}\M{M}^{-1}\M{B}\V{p}^* - \V{e} + \M{C}\V{u}^*.
  \label{eq:IPCSA_eqp}
\end{align}

The velocity $\V{u}$ can be recovered without solving a linear system, simply by substituting the pressure $\V{p}$ from the solution of \cref{eq:IPCSA_eqp} into \cref{eq:IPCSA_spliterr},
\begin{align}
  \V{u} = - \M{M}^{-1}\M{B}(\V{p} - \V{p}^*).
  \label{eq:IPCSA_velup}
\end{align}

The momentum-prediction \cref{eq:IPCSA_mom_with_guess} is solved followed by the pressure-correction \cref{eq:IPCSA_eqp} and the velocity update \cref{eq:IPCSA_velup} in an iterative manner until the required accuracy is reached. The resulting velocity field is projected into a BDM-type function space \citep{brezzi_mixed_1987} where it becomes exactly divergence free---the velocity flux is continuous across internal facets and the velocity field is pointwise divergence free inside each cell, see \citet{landet_pcorr_2019} for details. The result from this projection is stored both as $\vel^{n+1}$ and $\wvel^{n+1}$. The last step is slope limiting  $\vel^{n+1}$ such that any instabilities are prevented from growing through the time derivative.

Solving the coupled problem in \cref{eq:NS_mat_op} without some form of pressure and velocity splitting requires a direct solver, and such solvers scale badly in terms of parallel computing efficiency. The IPCS-A method can be solved using standard iterative Krylov methods which scale much better. Exact mass conservation, $\M{C}\V{u}-\V{e}=0$, is still ensured on the algebraic level in each iteration,
\begin{align}
  \M{C}\V{u} - \V{e}&=\  \M{C}\V{u} - \M{C}\V{u}^* + \M{C}\V{u}^* - \V{e} &
  \\ \notag
  & \overset{\mathclap{\text{\eqref{eq:IPCSA_spliterr}}}}=\ \ \M{C}\M{M}^{-1}\M{B}(\V{p} - \V{p}^*) + \M{C}\V{u}^* - \V{e}
  \\ \notag
  & \overset{\mathclap{\text{\eqref{eq:IPCSA_eqp}}}}=\ \ 0.
\end{align}

For the example simulations described in \cref{sec:num}, the maximum cell wise divergence error in the convecting velocity range from $10^{-5}$ to $10^{-3}$, while the convected velocity (which is slope limited) has a maximum cell wise divergence error ranging from $10^{2}$ to $10^{4}$. This is with $10^{-8}$ relative error and $10^{-10}$ absolute error convergence criteria in the pressure-correction Krylov solver. The result is that the simulation in \cref{sec:dambreak}, a dam breaking in a closed box, has less than \SI{3e-5}{\percent} change in total mass from time step 3 to the final time step (3415 steps, two seconds simulated time).

\section{Incoming waves and boundary reflections}
\label{sec:waves}

In our second numerical example, shown in detail in \cref{sec:cylinder}, we will study a surface-piercing vertical cylinder in an infinite wave field. A truncated computational domain is inevitably required to compute the solution in a finite amount of time. Our solution to the problem of reflected waves from the boundaries is to use a forcing-zone approach to wave damping \citep{peric_reliable_2016,peric_analytical_2018}. We impose Dirichlet boundary conditions to the velocity and density fields at both inlet and outlet boundaries. These boundaries are then padded by forcing zones inside the domain which penalise deviations from the undisturbed wave field. This damps out any free-surface disturbances caused by the structure without having to damp out the incident wave field itself. See \citet{peric_analytical_2018} for estimates of the minimum forcing zone size and the penalty magnitude needed to obtain a given reduction in reflected wave amplitudes.

\begin{figure}[ht]
  \centering
  \includegraphics[width=0.8\textwidth]{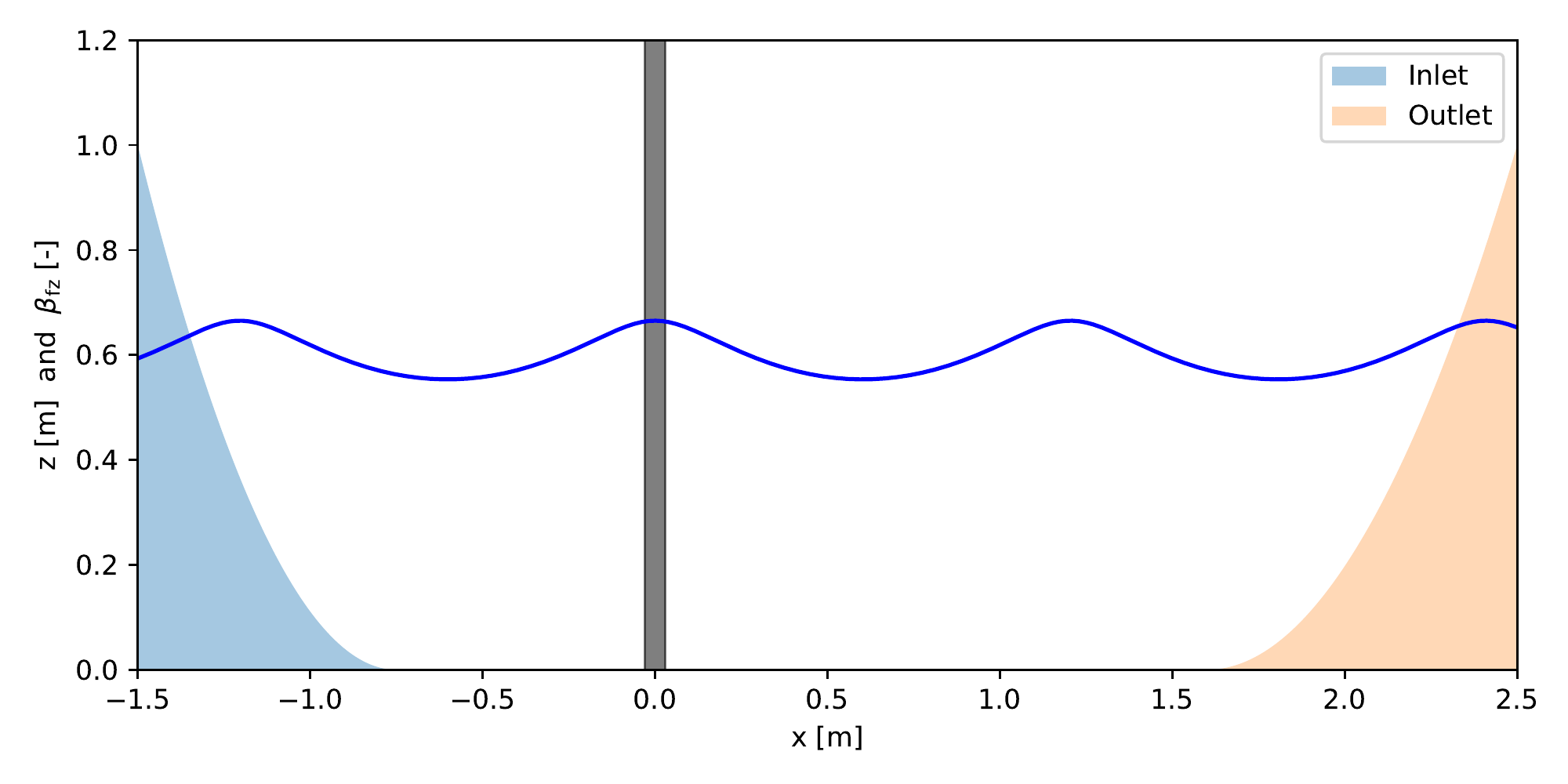}
  \caption{Forcing zones in the ``Cylinder in regular waves'' test case.}
  \label{fig:cylinder_setup}
\end{figure}

\Cref{fig:cylinder_setup} shows the forcing zones used in the test case described in \cref{sec:cylinder}. The inlet zone is \SI{0.75}{\meter} long and the outlet zone is \SI{1.0}{\meter} long. The shape of the zone is a quadratic polynomial with maximum value \num{1.0} at the boundary and zero first-derivative towards the inner domain. To force the solution towards the incident wave field, a penalty term is added to the momentum equations,
\begin{align}
\int_\tess \kappa_\text{fz}\,\beta_\text{fz} (\vel - \vel_D) \cdot \vtest \dif\pos,
\end{align}
where $\kappa_\text{fz}$ is the penalty parameter, in our calculations \num{10}, $\beta_\text{fz}$ is the zone shape shown in \cref{fig:cylinder_setup} and $\vel_D(x,z,t)$ is the incident-wave velocity field used for initial and boundary conditions. The same forcing-zone approach is added to the density-transport equation for the colour function with the same forcing-zone shapes and the same penalty parameter, $\kappa_\text{fz}=10$.

The initial and boundary conditions for the velocity and density fields are computed from Fenton stream function wave theory. This is a high-order regular wave theory based on approximating a stream function by a truncated Fourier series. This method of constructing non-linear regular waves was pioneered by \citet{dean_1965}. Our implementation is based on \citet{rienecker_fenton_1981}, which is often referred to as ``Fenton'' stream function wave theory to differentiate it from the original ``Dean'' stream function wave theory.

The Fenton method is based on collocation in $N+1$ points on the free surface along half the wave length, $\lambda$. The Fenton stream function, 
\begin{align}
  \Psi_w(x, z, t) = B_0 z + \sum_{j=1}^{N}B_j\frac{\sinh jkz}{\cosh jkD}\cos jkx,
  \label{eq:fenton_sf}
\end{align}
is non-linear in the wave height $\eta$ since $z=\eta$ in the collocation points. $\Psi_w$ does a-priori satisfy the bottom boundary condition at $z=0$ and also the Laplace equation $\nabla^2\Psi_w=0$. The following conditions are imposed in the Newton–Raphson iterations that are applied to compute the unknown coefficients in \cref{eq:fenton_sf}:
(i) the free surface is a stream line, $\Psi_w=$~const.,
(ii) the pressure is constant at the free surface,
(iii) the wave height is $H$, such that $\eta(0) - \eta(\lambda/2) = H$,
(iv) the mean wave elevation is $D$, such that $\int_0^{\lambda/2} \eta\dif x = D \lambda / 2$. See \cite{rienecker_fenton_1981} for details.

We use the same approach to find a compatible stream function for the air phase, $\Psi_a$, when $\Psi_w$ has been determined. The stream function in \cref{eq:fenton_sf} is now linear in the unknowns since $\eta$ is known, so the expansion coefficients can be found by a single linear solve. This gives two separate domains where the boundary conditions are satisfied and the divergence is zero, but the velocity parallel to the free surface has a discontinuity at the free surface since the method is based on potential theory and does not contain the viscosity that causes a velocity shear at the free surface. To approximate this, a blended stream function is constructed,
\begin{align}
  \Psi &= [1 - f(Z)] \Psi_{w}(x,z) + f(Z) \Psi_{a}(x,z), \\
  Z &= \frac{z - \eta(x)}{d - \eta(x)}\,, \notag
\end{align}
where the blending function $f(Z)$ is zero in the water and unity in the air above the blending zone. All blending is performed in the air phase, and the blending zone has height $d$, starting from the free surface. A fifth-order polynomial smooth step function is used for $f(Z)$. This function has zero first and second derivatives at the top and bottom of the blending zone. The resulting velocity field,
\begin{align}
  \mathbf{u}_x &= \ \ (1 - f) \frac{\partial\Psi_{w}}{\partial z} +
                  f \frac{\partial\Psi_{a}}{\partial z} -
                  \frac{d f}{d z}\Psi_{w}(x,z) +
                  \frac{d f}{d z}\Psi_{a}(x,z), \\
  \mathbf{u}_z &= -(1 - f) \frac{\partial\Psi_{w}}{\partial x} -
                  f \frac{\partial\Psi_{a}}{\partial x} +
                  \frac{d f}{d x}\Psi_{w}(x,z) -
                  \frac{d f}{d x}\Psi_{a}(x,z), \notag
\end{align}
is continuous and satisfies the continuity equation everywhere, the dynamic and kinematic free-surface boundary conditions at the interface, and the free-slip boundary conditions at the top and bottom of the domain. These properties make the blended solution a good choice to use as initial and boundary conditions in an exactly divergence-free Navier--Stokes solver.

\section{Implementation}
\label{sec:implementation}

The Ocellaris solver \citep{landet_joss_2019} is built on FEniCS \citep{logg_automated_2012} and PETSc \citep{petscAutoGen_davis2004algorithm,petscAutoGen_petsc-efficient,petscAutoGen_petsc-user-ref,petscAutoGen_Dalcin2011,petscAutoGen_hypre-web-page}. The overwhelming majority of the code, including the definition of the weak form and the time-stepping procedure, is implemented in Python 3. The FEniCS form compiler, FFC, is used to compile the weak form, defined in UFL Python format, to optimised C++ code \cite{kirby_compiler_2006,olgaard_automated_2008,alnaes_unified_2014}. Wherever tight loops over the mesh cells or facets are needed in other parts of the program, the loop is written in C++. For the simulation examples shown in \cref{sec:num}, the additional C++ code is restricted to parts of the VOF implementation and the velocity slope limiter.

Ocellaris is developed using automated unit and MMS (method of manufactured solutions) testing. Unfortunately, MMS testing of two-phase flows with discontinuous density fields is not well developed, so testing of this functionality involves running full time simulations, which is not done automatically on each change of the code as this would be too expensive. The Ocellaris program design is made up of independent pluggable components, letting the user define which combination of pressure-splitting scheme, free-surface model, slope limiter and more to use, simply by referencing the relevant components in the input file. This also means that there are automated MMS test of the implemented pressure-correction solvers, but they are tested with a single-phase flow model where classical analytical solutions are available, such as the Taylor--Green vortex \citep{green_mechanism_1937}.

\subsection{Example input file}
\label{sec:inpfile}

The following file listings show excerpts from the input file used to simulate regular waves passing a vertical, surface-piercing cylinder. More information about this example can be found in \cref{sec:cylinder}. Complete documentation of the input file format can be found in \citep{landet_ocellaris_user_guide_2019}, and the full input file can be found in \citep{landet_p3_zenodo_2019}. All Ocellaris input files are written on YAML format and start with a mandatory header section followed by an optional metadata section. An example of these two input file sections is shown in \cref{lst:cyl_meta}.

\begin{lstlisting}[language=yaml,label=lst:cyl_meta,caption={Input file header and metadata.}, captionpos=b]
ocellaris:
  type: input
  version: 1.0
metadata:
  author: Tormod Landet
  date: 2018-12-06
  description: |
      Surface piercing cylinder with regular waves
\end{lstlisting}
 
Ocellaris supports defining constants to be used throughout the input file. Defining these once on top of the file makes parameter studies and input changes easier to perform and makes the file easier to read. The definition of convenience constants and physical constants can be seen in \cref{lst:cyl_constants}.

\begin{lstlisting}[language=yaml,label=lst:cyl_constants,caption={Input file constants and physical constants.}, captionpos=b]
user_code:
    constants:
        H: 1.20      # Domain depth
        R: 0.03      # Cylinder radius
        L: 4.00      # Domain length
        B: 0.50      # Domain breadth
        C: 1.50      # Dist. from cylinder (origin) to inlet
        d: 0.60      # Water depth
        w: 0.75      # Length of the forcing zone
        wplus: 0.15  # Additional forcing zone at outlet
physical_properties:
    rho0: 1000.0
    nu0: 1.0e-6
    rho1: 1.0
    nu1: 1.5e-5
    g: [0, 0, -9.81]
\end{lstlisting}

The mesh section is shown in \cref{lst:cyl_msh}. The mesh file, created in gmsh \citep{gmsh09}, is loaded using the \texttt{meshio} Python package which implements readers and writers for many unstructured mesh file formats.
\begin{lstlisting}[language=yaml,label=lst:cyl_msh,caption={Input file mesh definition.}, captionpos=b]
  mesh:
      type: meshio
      mesh_file: ../meshA/cylinder.msh
      meshio_type: gmsh
\end{lstlisting}

Known fields functions are defined in \cref{lst:cyl_fields_wave,lst:cyl_fields_zone}. Known fields are used in Ocellaris to define initial and boundary conditions and also to define the location of the free-surface wave damping zones described in \cref{sec:waves}. The waves are defined using the \texttt{raschii} Python package which produces C++ code for the Fenton and Stokes regular wave models for use in FEniCS-based solvers. Note also that Ocellaris accepts Python expressions such as \texttt{"py\$ H - d"} instead of scalars, booleans and strings. These expressions can be used to define parameters in terms of the user-defined constants.
\begin{lstlisting}[language=yaml,label=lst:cyl_fields_wave,caption={Input file definition of the incoming wave field.}, captionpos=b]
fields:
-   name: waves
    type: RaschiiWaves
    wave_model: Fenton
    air_model: FentonAir
    model_order: 10
    still_water_position: py$ d
    depth: py$ d
    depth_above: py$ H - d
    blending_height: 0.3
    wave_height: 0.11187
    wave_length: 1.20444
\end{lstlisting}

The damping zone in \cref{lst:cyl_fields_zone} is implemented as a generic scalar field. The C++ code is compiled inside a namespace that includes all user-specified constants and an array, \texttt{x}, defining the coordinates where the field is to be evaluated. Using C++ lambdas allows using multi-line expressions to compute the field value. Standard C++ expressions, such as \texttt{"x[0] + x[1]*x[2]"} ($x + yz$), can also be given for fields which do not need multiple statements to compute the field value. The inlet damping zone is defined equivalently to the outlet damping zone, but is not shown here for brevity.
\begin{lstlisting}[language=yaml,label=lst:cyl_fields_zone,caption={Input file definition of the outlet damping zone location.}, captionpos=b]
-   name: outlet zone
    type: ScalarField
    variable_name: beta
    stationary: yes
    cpp_code: |
        [&]() {
            double dz0 = (L - (w + wplus)) - C;
            double dz1 = (L - 0) - C;
            if (x[0] < dz0) {
                return 0.0;
            } else if (x[0] > dz1) {
                return 1.0;
            } else {
                return pow((x[0] - dz0)/(dz1 - dz0), 2);
            }
        }()
\end{lstlisting}

A definition of a forcing zone is shown in \cref{lst:cyl_fz}. There are four such zones in the simulation, damping the momentum and density fields at the inlet and outlet, see \cref{sec:waves}. The zone in the example shows the damping of the momentum equations at the outlet boundary. It uses the previously defined known fields \texttt{waves} and \texttt{outlet zone} to specify the target value and field location.
\begin{lstlisting}[language=yaml,label=lst:cyl_fz,caption={Input file definition of a momentum damping zone.}, captionpos=b]
forcing_zones:
-   name: outlet velocity damping
    type: MomentumForcing
    zone: outlet zone/beta
    target: waves/u
    penalty: 10
    plot: no
\end{lstlisting}

The initial conditions are defined in \cref{lst:cyl_ic}. The naming scheme uses the postfix \texttt{p} to specify the value of a field at $t=0$, the previous time-step value, and \texttt{0} and \texttt{2} to specify the $x$ and $z$ directions respectively. To use higher-order time stepping from the start, the values at $t=-\Delta t$ can be specified by using the \texttt{pp} prefix, but that is normally only done in convergence tests where the analytical solution is known.
\begin{lstlisting}[language=yaml,label=lst:cyl_ic,caption={Input file definition of initial conditions.}, captionpos=b]
initial_conditions:
    cp:  # c is the VOF colour function
        function: waves/c
    up0:
        function: waves/uhoriz
    up2:
        function: waves/uvert
\end{lstlisting}

\Cref{lst:cyl_bc} shows the definition of boundary conditions for the inlet. The inside code is used to select the facets on the inlet. If a boundary region is marked with an integer identifier in the mesh generator, then facet selection can be based on this identifier instead. For the cylinder-in-waves example the boundary facets are all marked with C++ code as shown in \cref{lst:cyl_bc}. The \texttt{on\_boundary} boolean flag is \texttt{true} for external facing facets.
\begin{lstlisting}[language=yaml,label=lst:cyl_bc,caption={Input file definition of boundary conditions.}, captionpos=b]
boundary_conditions:
-   name: Inlet
    selector: code
    inside_code: "on_boundary and x[0] < 0 - C + 1e-5"
    u0:
        type: FieldFunction
        function: waves/uhoriz
    u1:
        type: ConstantValue
        value: 0
    u2:
        type: FieldFunction
        function: waves/uvert
    c:
        type: FieldFunction
        function: waves/c
\end{lstlisting}

The Navier--Stokes solver section specifies which velocity--pressure splitting scheme to use, the iteration tolerances and the PETSc Krylov solver parameters. Any PETSc configuration variable can be changed for maximum flexibility. In the example shown in \cref{lst:cyl_solver}, the Ocellaris default options for the momentum and pressure PETSc solvers are used and only the convergence criteria and the number of inner iterations (the number of pressure corrections per time step) are changed. After ten time steps, only two pressure corrections are performed per time step. The Krylov solver tolerances are given as three numbers; the tolerance for the first three pressure corrections, the value for the mid range of pressure corrections, and finally the values for the last five pressure corrections. Since, after the first ten time steps, there are only two pressure corrections per time step,  only the last value in the lists matter. This gradual decrease of tolerances is done to avoid spending a lot of time in the Krylov solver in the beginning of the simulation when the pressure corrections are not converged and exact answers are not needed.

\begin{lstlisting}[language=yaml,label=lst:cyl_solver,caption={Input file configuration of the solver.}, captionpos=b]
solver:
  type: IPCS-A
  num_inner_iter: py$ 10 if it < 3 else (5 if it < 10 else 2)
  allowable_error_inner: 1.0e-4
  use_stress_divergence_form: yes
  u:
      inner_iter_control: [3, 5]
      inner_iter_rtol: [1.0e-2, 1.0e-4, 1.0e-6]
      inner_iter_atol: [1.0e-2, 1.0e-4, 1.0e-6]
      inner_iter_max_it: [50, 200, 9999]
  p:
      inner_iter_control: [3, 5]
      inner_iter_rtol: [1.0e-4, 1.0e-6, 1.0e-8]
      inner_iter_atol: [1.0e-6, 1.0e-8, 1.0e-10]
      inner_iter_max_it: [50, 200, 9999]
\end{lstlisting}

The multiphase VOF input sections are shown in \cref{lst:cyl_mps}. The HRIC VOF scheme is selected and 5 sub-cycles---the number of advection steps of the VOF colour function per time step of the Navier--Stokes solver---are applied to maximise sharpness of the interface by lowering the effective Courant number in the VOF solver.
\begin{lstlisting}[language=yaml,label=lst:cyl_mps,caption={Input file configuration of the multi-phase solver.}, captionpos=b]
multiphase_solver:
    type: BlendedAlgebraicVOF
    num_subcycles: 5
convection:
    c:
        convection_scheme: HRIC
\end{lstlisting}

The final excerpt from the input file is shown in \cref{lst:cyl_slopelim}. Here the velocity slope limiter is configured to use the scalar \texttt{HierarchicalTaylor} slope-limiting method by \citet{kuzmin_vertex-based_2010} for each component of the convected velocity field.

\begin{lstlisting}[language=yaml,label=lst:cyl_slopelim,caption={Input file slope-limiter configuration.}, captionpos=b]
slope_limiter:
  u:
      method: Componentwise
      comp_method: HierarchicalTaylor
\end{lstlisting}

\section{Numerical examples}
\label{sec:num}

The following example simulations have been run in Ocellaris. Source code and input files, which can be used to reproduce all the results, can be found in \citep{landet_p3_zenodo_2019}. Visualisations and isosurfaces have been made in Paraview \citep{paraview_2005}. The meshes, all available for download, have been generated and optimised in Gmsh \citep{gmsh09}.

\subsection{3D dam breaking}
\label{sec:dambreak}

\citet{kleefsman_vof_2005} presents experimental and numerical results for a 3D dam-breaking test case where a small container is subjected to a fast-moving front of water that splashes over and around the container. The tank is \SI{3 x 1 x 1}{\metre} and the container is \SI{16 x 40 x 16}{\centi\metre}. There are three surface-height probes that are initially dry and one that is in the fluid (H4). The container is fitted with eight pressure gauges close to the centre line. The locations of the gauges can be found in \citet{issa_ercoftac_2006}  along with an exact description of the tank and container geometries.

\begin{figure}[ht]
  \centering
  \includegraphics[width=0.8\textwidth]{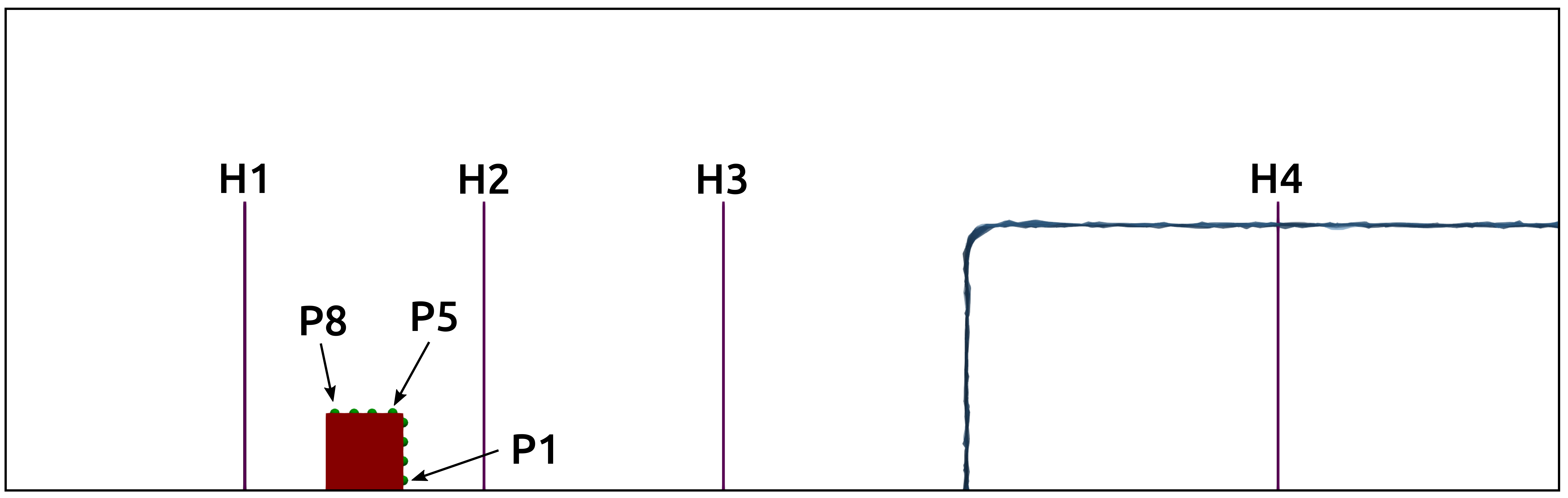}
  \caption{3D dam breaking. Sketch of the initial conditions along with the location of the surface-height probes H1--H4 and pressure gauges P1--P8.}
  \label{fig:green_water_setup}
\end{figure}

\begin{figure}[htb]
  \centering
  \includegraphics[width=0.95\textwidth]{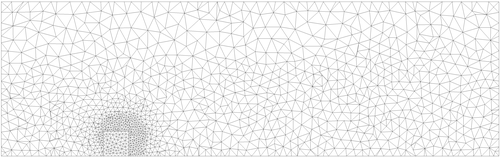}
  \caption{Longitudinal cut through the centre of the medium-density mesh. The mesh does not conform to the cut plane, making some elements look skewed.}
  \label{fig:greenwater_mesh}
\end{figure}

The Ocellaris simulations have been run based on irregular tetrahedral meshes. The same mesh input has been used with three different mesh densities, resulting in a coarse mesh with a total of \num{18676} tetrahedra, a medium-density mesh with \num{40450} tetrahedra and a fine mesh with \num{123628} tetrahedra. A cross-sectional view of the medium-density mesh can be seen in \cref{fig:greenwater_mesh}. The time step is adaptively controlled based on the maximum cell-based \eqref{eq:Co} and facet-based \eqref{eq:Cof} Courant numbers. The time step is halved if one of the two is above \num{0.3} and doubled if they are both below \num{0.05}. The Courant numbers are computed for each cell and facet as
\begin{align}
\Co &= \frac{|\vel|\Delta t}{D_c} \qquad \text{(cell average)} \label{eq:Co}\\
\Co_f &= \frac{\vel\cdot\normal\,S_f\Delta t}{V_c N_\text{sc}}  \qquad \text{(facet average)} \label{eq:Cof}
\end{align}
where $D_c$ is the cell diameter, $S_f$ is the facet area, and $V_c$ is the cell volume. The facet-based Courant number is used in the HRIC transport scheme for the colour function and this is sub-cycled with $N_\text{sc}$ sub-cycles per time step. We have used $N_\text{sc}=5$ which makes the cell- and facet-based Courant numbers similar in magnitude.

\begin{figure}[htb]
  \centering
  \includegraphics[width=0.45\textwidth]{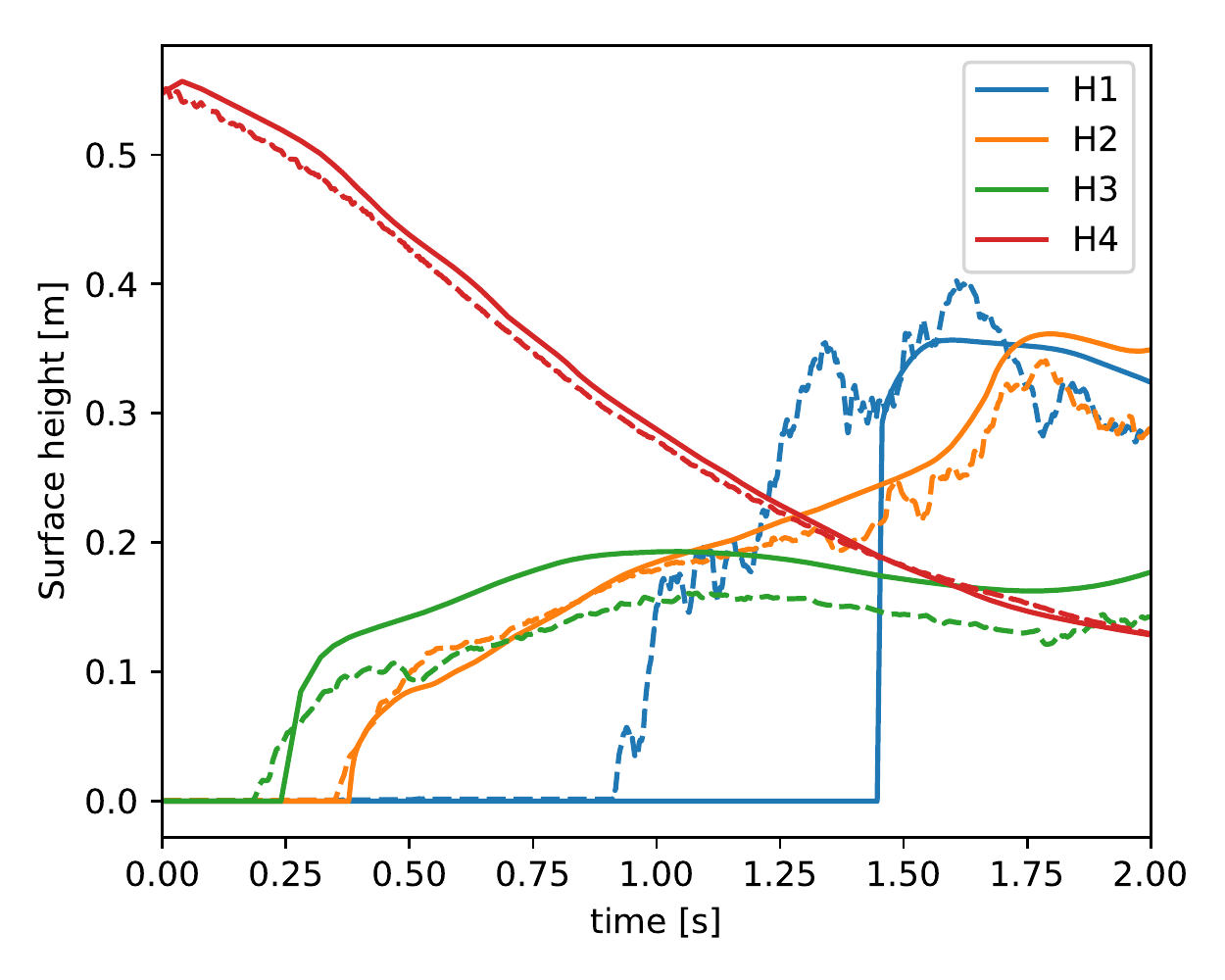} \hspace{6mm}
  \includegraphics[width=0.45\textwidth]{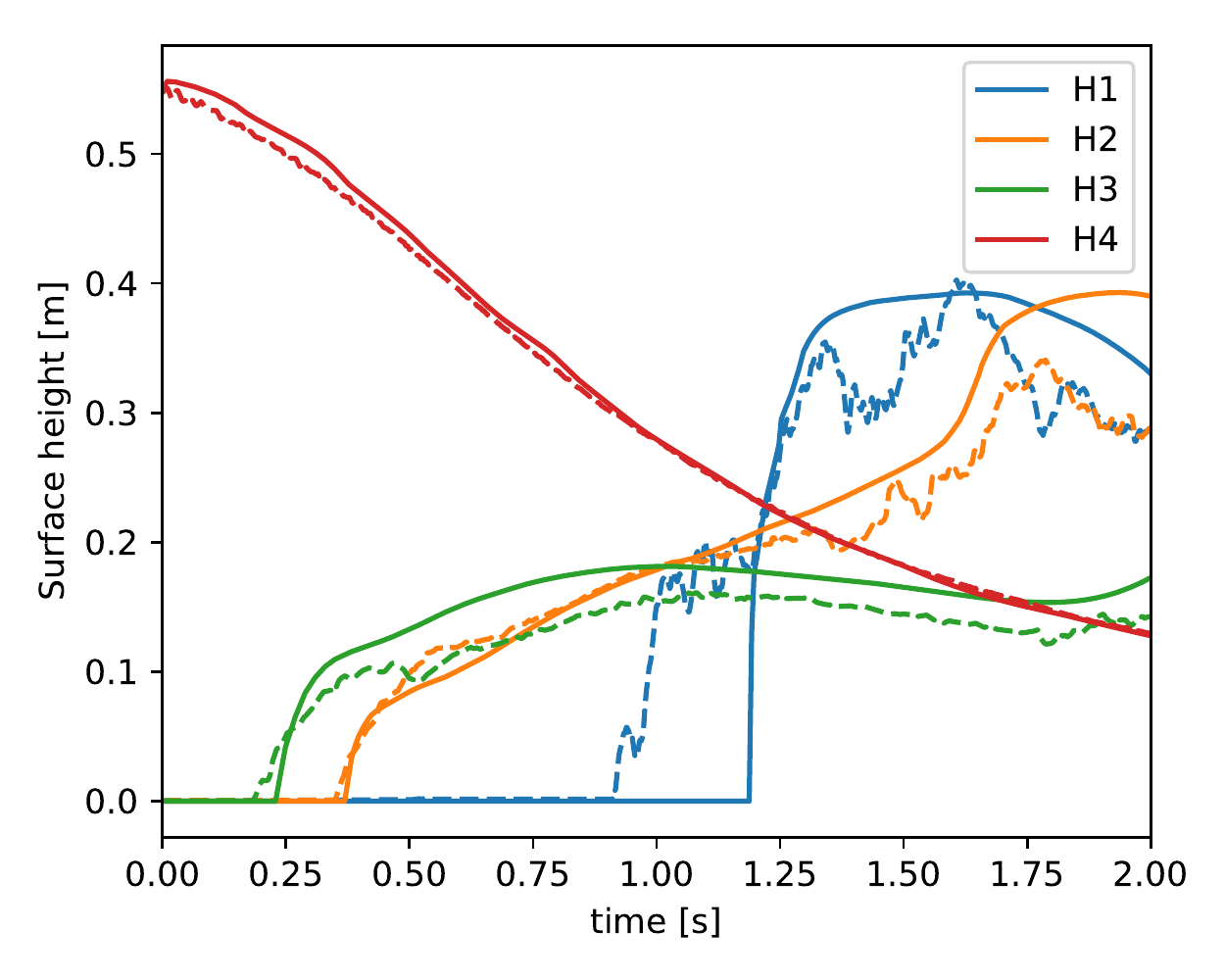}
  \caption{Free-surface probes. Medium mesh (left) and fine mesh (right). Experimental data shown with dashed lines.}
  \label{fig:greenwater_free_surface}
\end{figure}

\begin{figure}[htb]
  \centering
  \begin{minipage}[t]{.46\textwidth}
    \centering
    \includegraphics[width=0.95\textwidth]{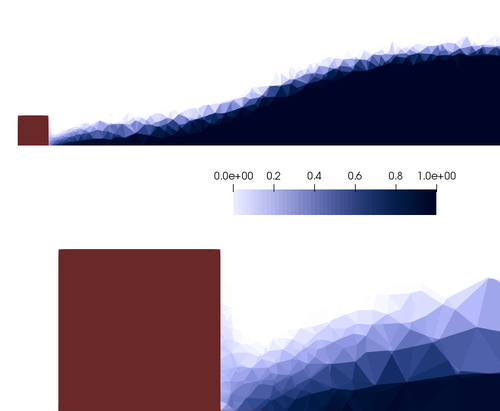}
    \caption{Fine mesh, $t=\SI{0.4075}{\second}$. Centered slice with closeup of the colour function near the container.}
    \label{fig:greenwater_diffusion}
  \end{minipage}
  \hspace{0.05\textwidth}
  \begin{minipage}[t]{.46\textwidth}
    \centering
    \includegraphics[width=0.95\textwidth]{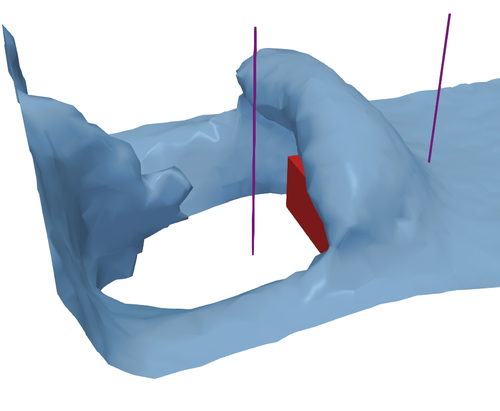}
    \caption{Fine mesh, $t=\SI{1.1075}{\second}$. $\cfunc=\num{0.5}$ isosurface from Paraview.}
    \label{fig:greenwater_splash}
  \end{minipage}
\end{figure}

\Cref{fig:greenwater_free_surface} shows the evolution of the four free-surface probes for the medium and fine mesh. The fine mesh fits best with the experiments, but both mesh resolutions show good comparison for probes H2--H4. The H1 probe, where the free-surface height is multi valued, does not compare well. It is not clear exactly what heigh is measured in the experiments at this location, we have reported the topmost surface intersection. A visualisation of the colour function at $t=\SI{0.4075}{\second}$, when pressure probe P2 spikes in the experimental results, is shown in \cref{fig:greenwater_diffusion}. The colour function isosurface from $t=\SI{1.1075}{\second}$, when the free surface is multi-valued behind the container, can be seen in \cref{fig:greenwater_splash}.

\begin{figure}[htb]
  \centering
  \includegraphics[width=0.45\textwidth]{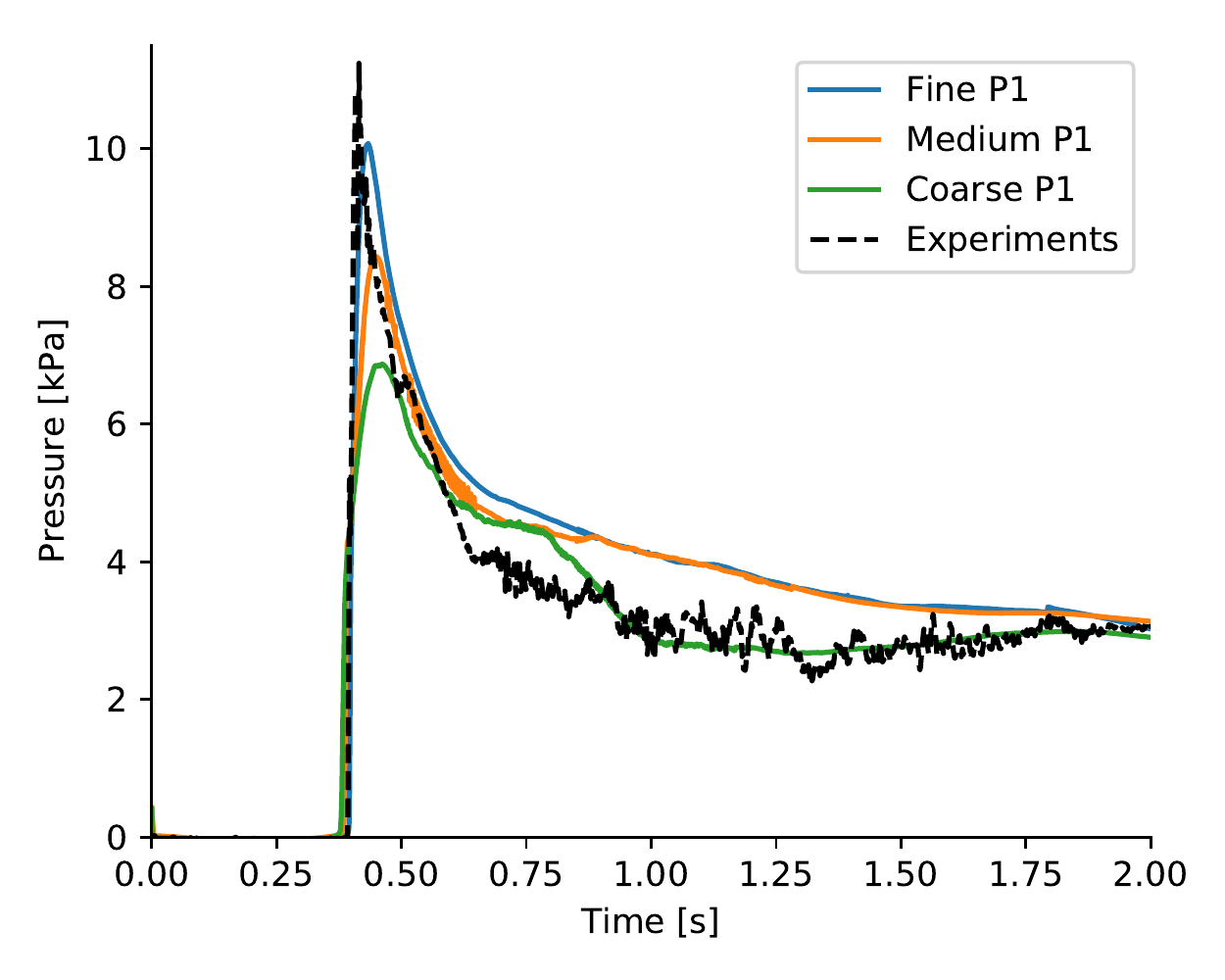} \hspace{5mm}
  \includegraphics[width=0.45\textwidth]{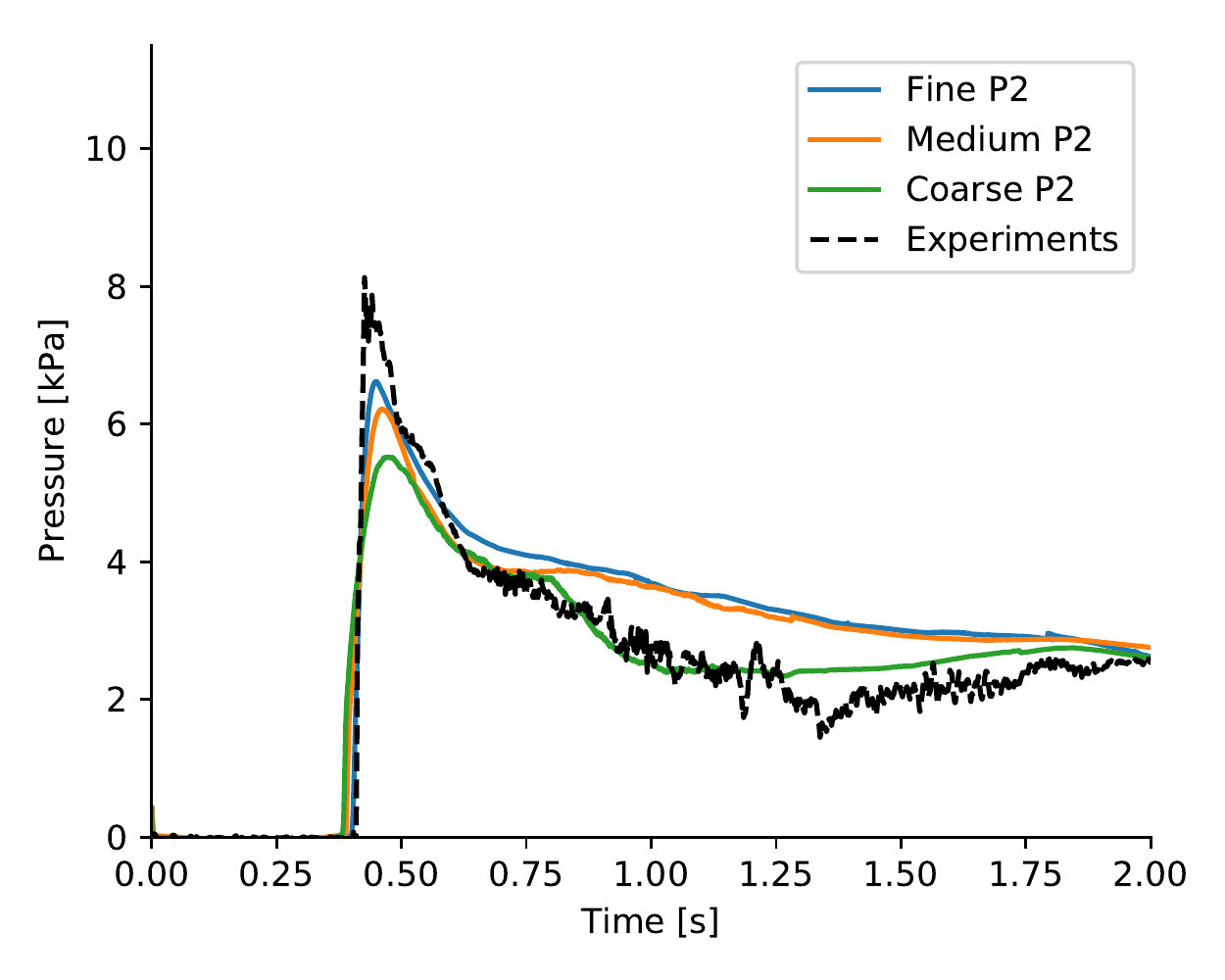}
  \includegraphics[width=0.45\textwidth]{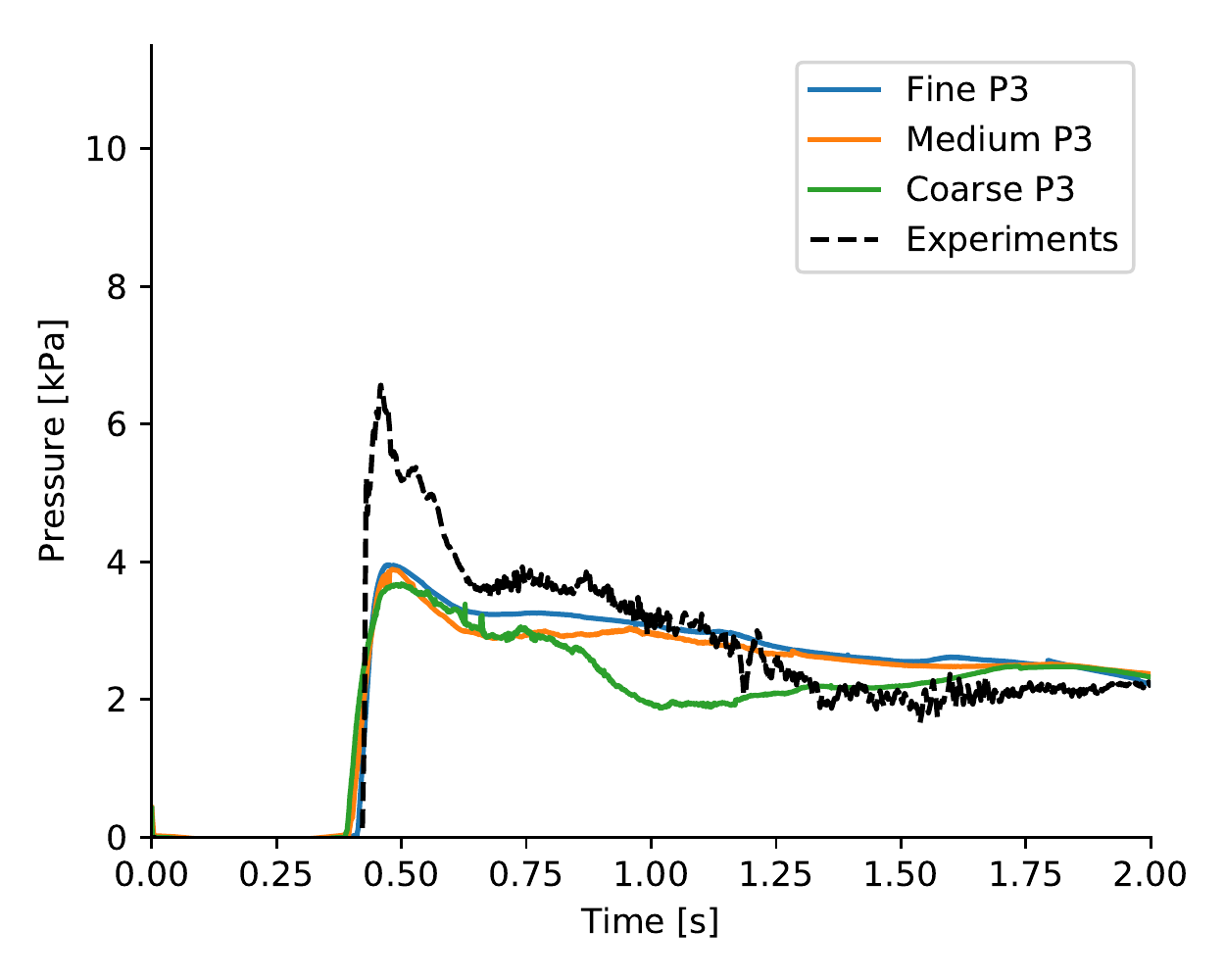} \hspace{5mm}
  \includegraphics[width=0.45\textwidth]{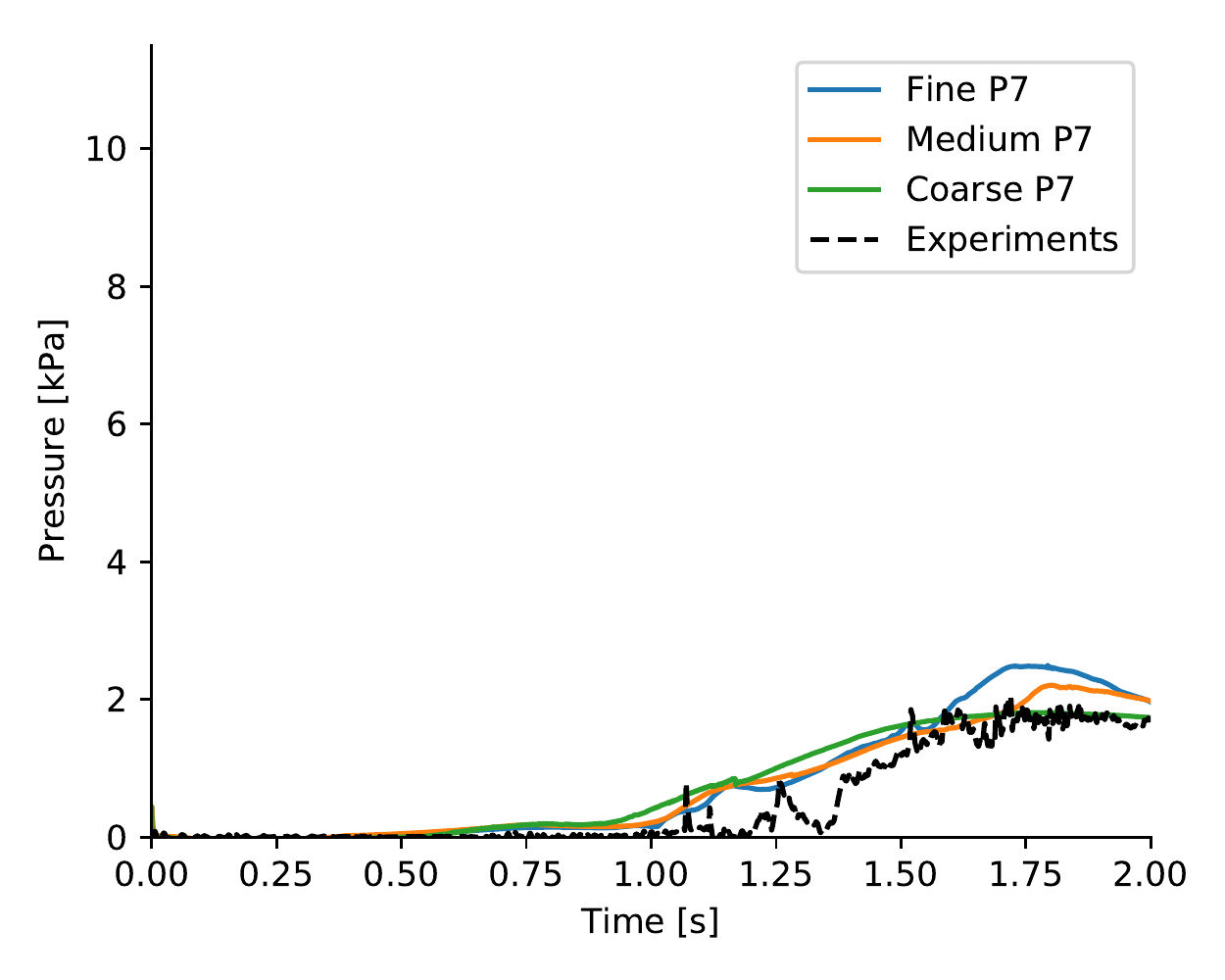}
  \caption{3D dam breaking, pressure probe results for all three meshes compared to the time series from the experiments.}
  \label{fig:greenwater_pressure}
\end{figure}

The pressure probe time series are shown in \cref{fig:greenwater_pressure}. The pressure probes on top of the container, P5--P8, all show very similar behaviour, we have included the results for probe P7 as an example, the other plots would have been roughly identical. On the side of the container facing the wave, the mesh convergence can be seen clearly in the P1 and P2 probes. Further up the container wall the pressure peak is less pronounced in the numerical results, possibly due to the interface being smeared over more cells than optimal, creating a more gradual rise in density at the pressure sensors than what happened in the experiment.

\subsection{Cylinder in regular waves}
\label{sec:cylinder}

\citet{grue_huseby_2002} gives experimental results for the inline force on a cylinder in regular waves. Figure 3d in that reference is used by \citet{paulsen_forcing_2014} as a benchmark problem, and in this numerical example we will do the same. Fenton stream-function waves are applied in a numerical wave tank which is \SI{4}{\meter} long and \SI{0.5}{\meter} wide. Our numerical domain height is \SI{1.2}{\meter} meter with free-slip BCs at the top. The still-water depth is $h=\SI{0.6}{\meter}$, the wave heigh is $H=\SI{0.112}{\meter}$ and the wave length is $\lambda=\SI{1.204}{\meter}$, leading to a wave number of $k=\SI{5.216}{\per\meter}$, a wave period of $T=\SI{0.843}{\second}$ and a phase speed of $c=\SI{1.429}{\meter\per\second}$. The situation is shown in \cref{fig:cylinder_setup} with a surface-piercing vertical cylinder of radius $R=\SI{3}{\centi\meter}$. Nondimensional parameters are $kR=\num{0.16}$, $kh=\num{0.16}$ and $kH=\num{0.58}$. The order of the Fenton stream function wave in \cref{eq:fenton_sf} is $N=\num{10}$.

\begin{figure}[htb]
  \centering
  \includegraphics[width=0.95\textwidth]{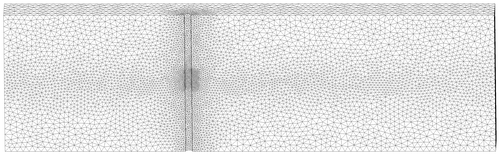}
  \caption{Mesh used for the cylinder-in-waves test case, only half the domain is included, symmetry is applied on the center face.}
  \label{fig:cylinder_mesh}
\end{figure}

The mesh is shown in \cref{fig:cylinder_mesh}. The mesh cells in the wave propagation zone have characteristic lengths of about \SI{2.5}{\centi\meter}, giving approximately 4.5 elements per wave height. To test wave propagation through the domain a similar mesh is used with the same refinement around the cylinder location, but without the cylinder being present. Symmetry boundary conditions are applied to the centre longitudinal plane shown prominently in \cref{fig:cylinder_mesh}. The inlet and outlet boundary conditions with associated forcing zones are implemented as described in \cref{sec:waves}. The cylinder, when it is included, is non-slip in the horizontal plane, but the vertical velocity component is allowed to slip to avoid the free surface sticking to the cylinder. This approximation is done since our mesh resolution is not fine enough to model the true wetting dynamic with appropriate slip lengths. The longitudinal wall away from the cylinder is modelled as free-slip boundary, the same is true for the top and bottom surfaces of the tank.

\begin{figure}[htb]
  \centering
  \includegraphics[width=0.40\textwidth]{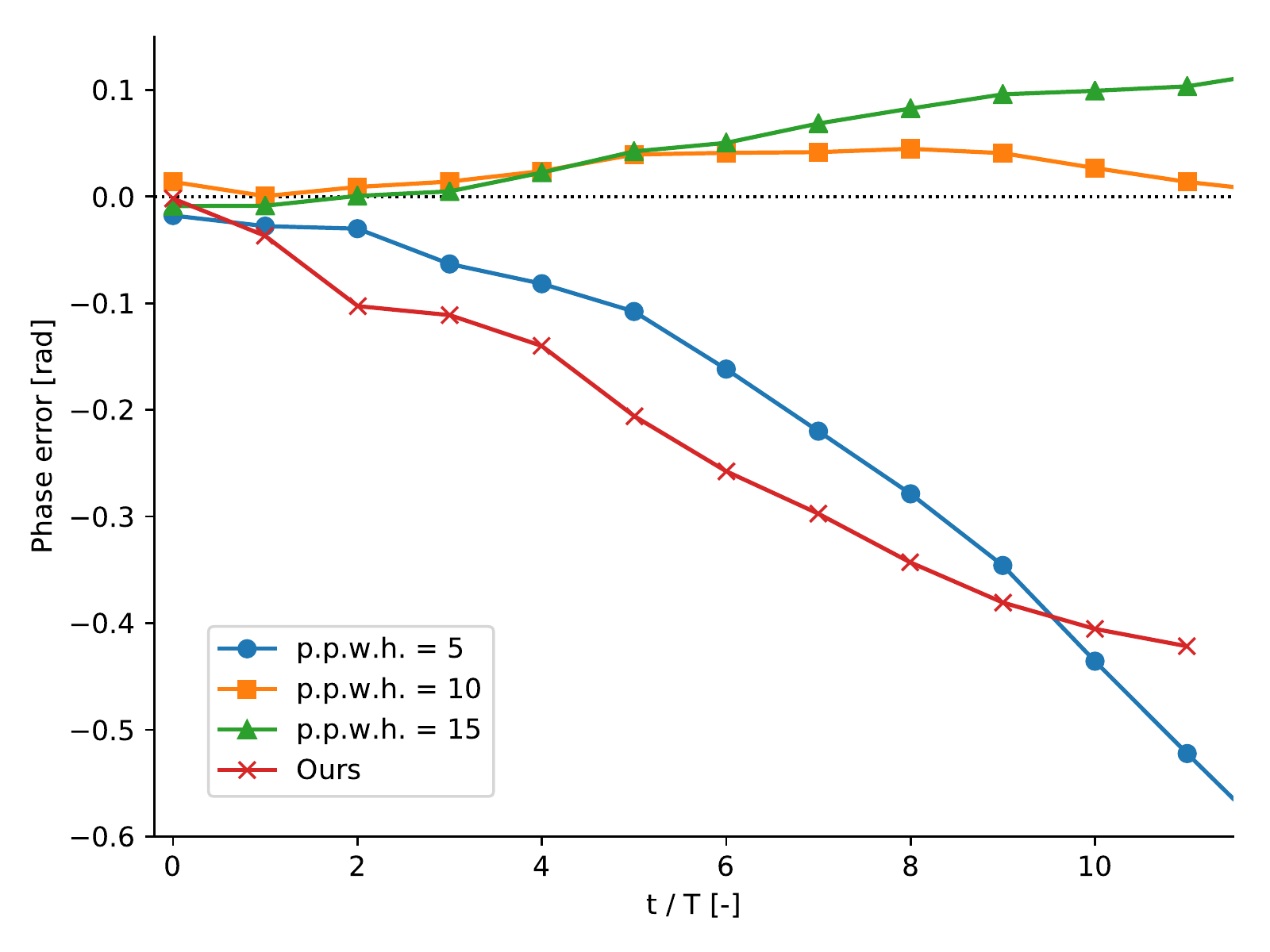} \hspace{1cm}
  \includegraphics[width=0.40\textwidth]{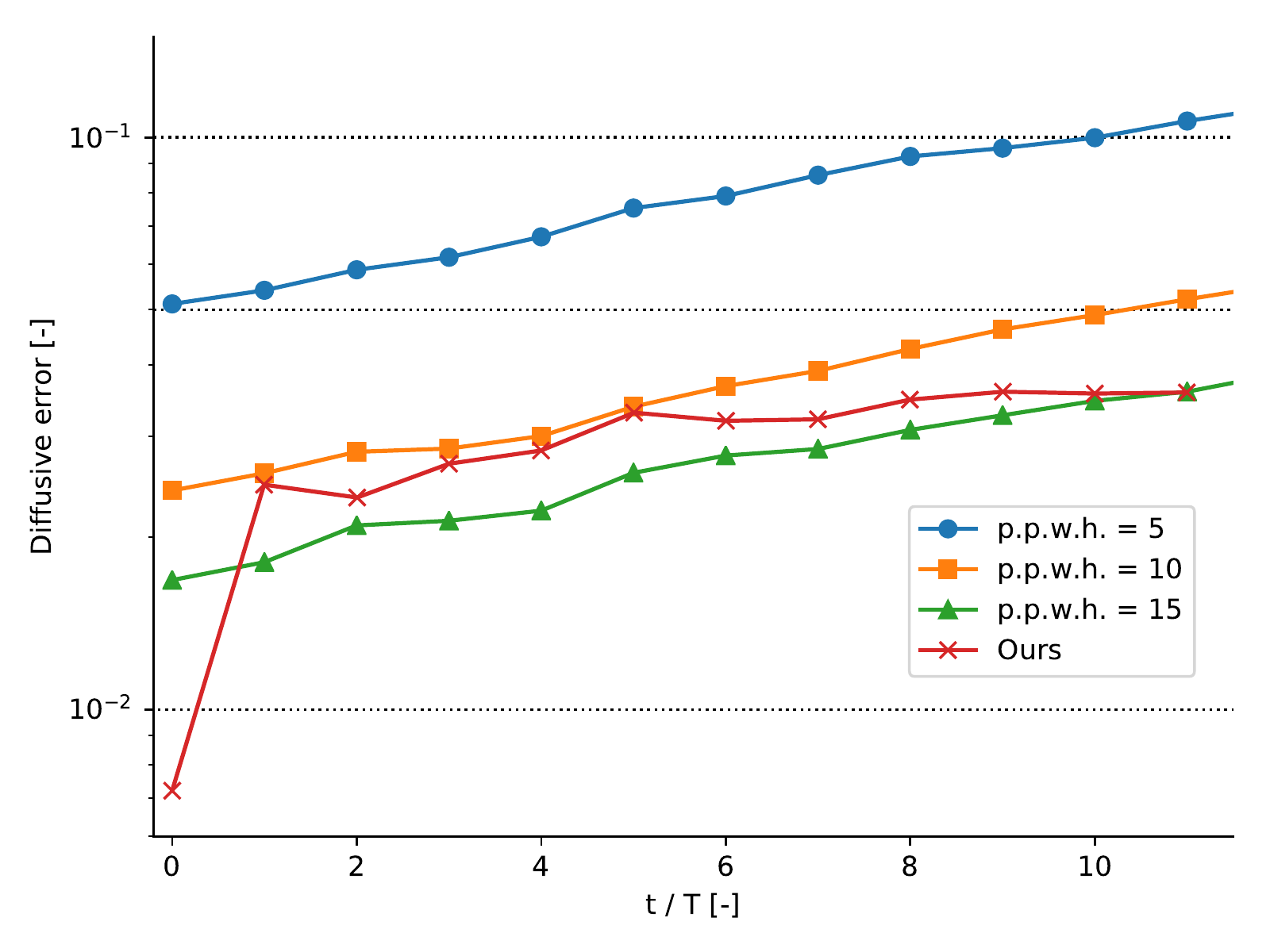}
  \caption{Phase error (left) and diffusive error (right). Results from \citet{paulsen_forcing_2014} where p.p.w.h. means mesh points per wave heigh. Our results have about 4.5 cells per wave height. $T$ is the wave period.} 
  \label{fig:wave_errors}
\end{figure}

The difference between the target wave elevation from the Fenton stream function and the VOF free surface is computed based on the intersection of the $\cfunc=0.5$ isosurface and a centred longitudinal plane through the numerical domain without the cylinder. The region from $\sfrac{2}{3}\,\lambda$ in front of the cylinder position to $\sfrac{1}{3}\,\lambda$ behind the cylinder position is used to compute the difference between the phase of the numerical free surface and the stream function's free surface. The diffusive error is computed in the same region as $\frac{1}{H} \sqrt{\langle\eta'^2\rangle}$ where $\eta'$ is the difference between the VOF and the stream-function wave elevation after correcting for the phase error and $\langle\cdot\rangle$ denotes the mean over about \num{300} points where $\eta'$ is sampled. \Cref{fig:wave_errors} shows the results compared to \citet[figures 2 and 3]{paulsen_forcing_2014}, but note that our results are from a 3D domain with forcing zones as described in \cref{sec:waves}, while their results are from 2D simulations in a periodic domain which will let errors develop more easily over time.

\begin{figure}[htb]
  \centering
  \includegraphics[width=1.0\textwidth]{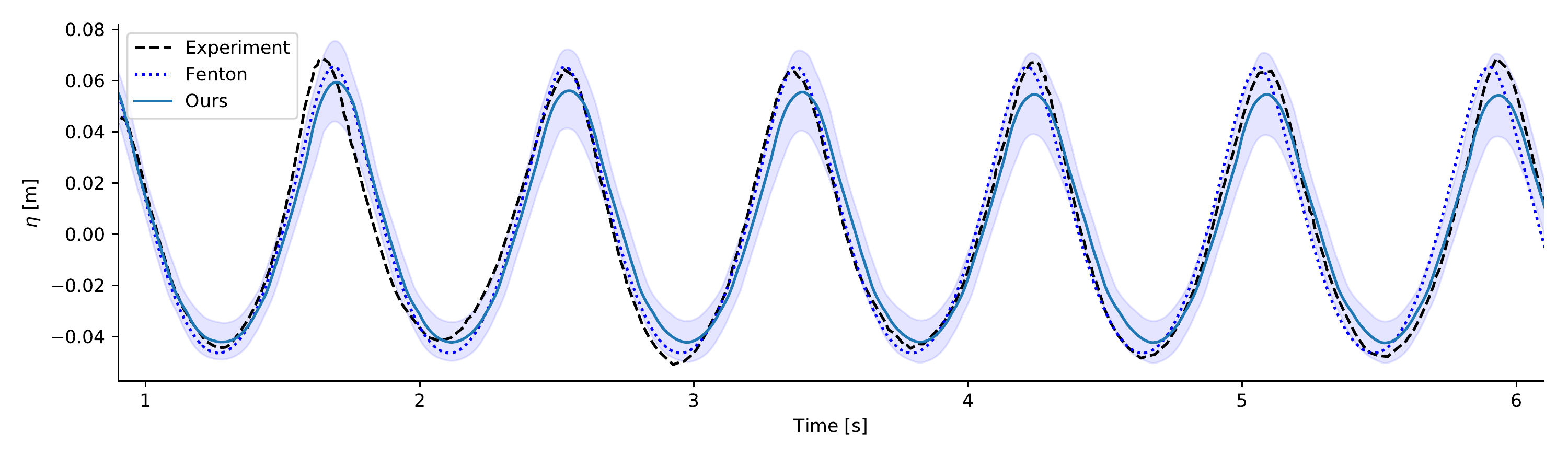}
  \caption{Surface elevation at the cylinder location, comparing laboratory experiment by \citet{grue_huseby_2002} to a Fenton stream-function wave and our numerical results. The $\cfunc=0.5$ isosurface is shown along with a shaded area showing $\cfunc\in[0.35, 0.65]$.}
  \label{fig:cylinder_wavecompare}
\end{figure}

\begin{figure}[htb]
  \centering
  \includegraphics[width=1.0\textwidth]{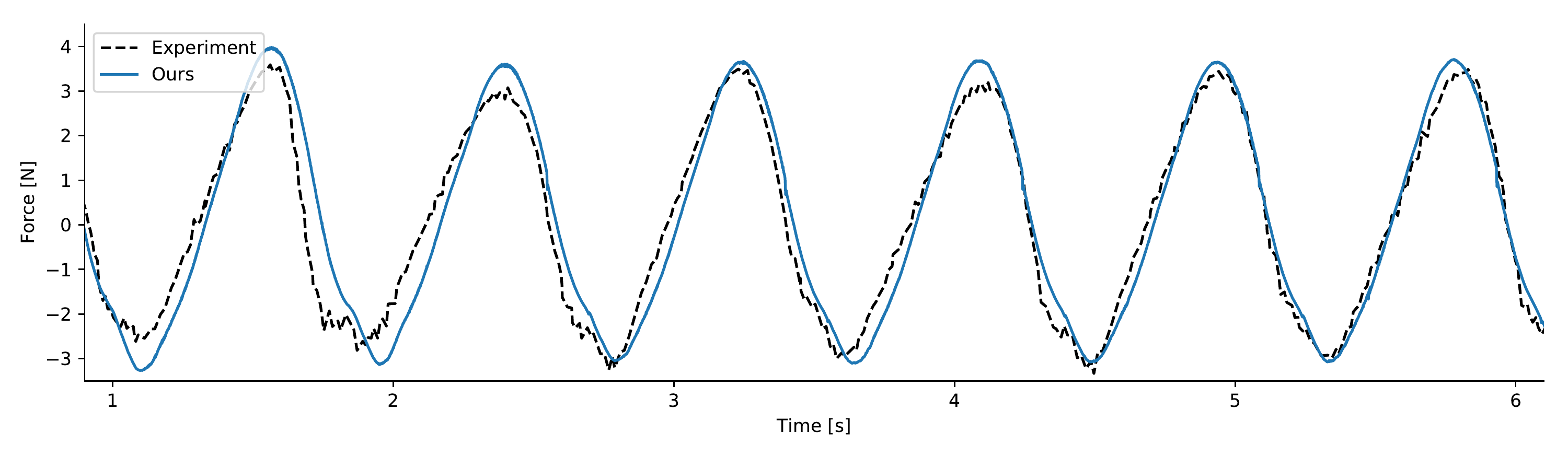}
  \caption{Inline force on cylinder. Our numerical results compared to the laboratory experiment by \citet{grue_huseby_2002}.}
  \label{fig:cylinder_forcecompare}
\end{figure}

\begin{figure}[htb]
  \centering
  \includegraphics[width=0.9\textwidth]{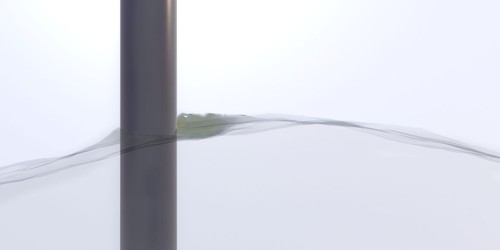}
  \caption{Snapshot of free surface slightly after a wave crest has passed from left to right. Isosurface made by Paraview and visualised in Blender.  The domain was mirrored about the symmetry plane when making the visualisation.}
  \label{fig:cylinder_ridge}
\end{figure}

Our numerical results are compared to the experimental results by \citet{grue_huseby_2002} in \cref{fig:cylinder_wavecompare,fig:cylinder_forcecompare}. The time series from figures 3c and 3d in their work have been shifted in time by \SI{-15.63}{\second} to align the wave phases. \Cref{fig:cylinder_wavecompare} shows the wave elevation and also includes the Fenton stream-function solution to show the phase and diffusive errors in a more intuitive way. This also confirms that the applied stream-function wave is similar to the wave obtained in the laboratory wave flume. The inline force shown in \cref{fig:cylinder_forcecompare} is computed as the total force on the cylinder in the longitudinal direction, positive towards the outlet. Note the bumps in the force signal right before each trough, the ``secondary load cycle''. The associated free-surface ridge behind the cylinder at this time is shown in \cref{fig:cylinder_ridge}.

\section{Discussion}
\label{sec:discussion}

Both test cases presented in \cref{sec:num} show that the Ocellaris DG FEM solver is able to simulate complex two-phase flows in 3D. The pressure probes in the dam breaking test show good agreement with the experiments. Pressure probe P1 matches very well and the convergence between the mesh resolutions is clear. The impact pressures higher up on the vertical front wall of the container are less peaked than the measurements. They are most likely softened by the diffusive free surface, which can be seen in \cref{fig:greenwater_diffusion}. The surface height probes match well, except for the probe behind the container where the flow is very complex and the free surface is multi valued. For the second test case the final comparison of the inline force on the cylinder shows an excellent match and the expected free-surface ridge behind the cylinder is observed.

The comparatively primitive free-surface capturing method is the weak point of the current implementation. Using a piecewise constant density field on the same mesh as the piecewise quadratic convecting velocity means that a lot of the data is thrown away when it comes to the advection of the free surface. Quantities directly related to the jump in the density field will always be limited to first-order convergence, but a more sophisticated free-surface capturing method could provide a sharper profile, hopefully eliminating the softening of the impact pressures in the first test case and removing much of the diffusive error in the wave propagation test. The excellent match of the inline force in the second test case should be treated with some scepticism considering the observed lowering of the peak height of the incident wave. The diffusive interface may have counteracted the expected lowering of the peak force by increasing the air phase pressures through an artificially high fluid density near the free surface. 

The reason we selected to use only 4.5 elements per wave height for the wave propagation test is that the velocity function space has \num{30} unknowns per mesh cell, so it is necessary to limit the number of cells more than in lower-order methods. The simulation has about \num{120000} cells and is run on \num{32} CPUs, so significantly increasing the mesh density will come at a large cost. This highlights the need for more research into free-surface capturing schemes that are designed specifically for high-order methods. 

The main advantage of the method is the potential for using fewer elements away from the free surface due to the quadratic approximating polynomials. The dependency on having a very high mesh resolution in the free-surface region must be removed if the method is to be a preferable solution for general free-surface simulations. Another approach would be to make the method $h$--$p$ adaptive.
A relevant issue for further research, which we have not touched on here, is that we observe that the number of iterations in the Krylov solver for the momentum equation increases a lot when the density ratio increases. The momentum equation can most likely be preconditioned or stabilised to stop this tendency.

We have shown that it is viable to use Ocellaris' exactly divergence-free high-order discontinuous Galerkin finite element method with velocity slope limiting to simulate realistic air/water free-surface physics, but some work remains to be fully competitive with established low-order methods.

\section*{Acknowledgements}

The simulations were performed on resources provided by UNINETT Sigma2, the National Infrastructure for High Performance Computing and 
Data Storage in Norway.

\bibliographystyle{plainnat} 
\bibliography{refs.bib}

\begin{thebibliography}{42}
\providecommand{\natexlab}[1]{#1}
\providecommand{\url}[1]{\texttt{#1}}
\expandafter\ifx\csname urlstyle\endcsname\relax
  \providecommand{\doi}[1]{doi: #1}\else
  \providecommand{\doi}{doi: \begingroup \urlstyle{rm}\Url}\fi

\bibitem[Ahrens et~al.(2005)Ahrens, Geveci, and Law]{paraview_2005}
James Ahrens, Berk Geveci, and Charles Law.
\newblock {ParaView}: An end-user tool for large-data visualization.
\newblock In Charles~D. Hansen and Chris~R. Johnson, editors,
  \emph{Visualization Handbook}, pages 717--731. Butterworth-Heinemann, 2005.

\bibitem[Alnæs et~al.(2014)Alnæs, Logg, Ølgaard, Rognes, and
  Wells]{alnaes_unified_2014}
Martin~S. Alnæs, Anders Logg, Kristian~B. Ølgaard, Marie~E. Rognes, and
  Garth~N. Wells.
\newblock Unified {Form} {Language}: {A} {Domain}-specific {Language} for
  {Weak} {Formulations} of {Partial} {Differential} {Equations}.
\newblock \emph{ACM Trans. Math. Softw.}, 40\penalty0 (2):\penalty0 9:1--9:37,
  March 2014.
\newblock ISSN 0098-3500.

\bibitem[Arnold(1982)]{arnold_interior_1982}
Douglas~Norman Arnold.
\newblock An interior penalty finite element method with discontinuous
  elements.
\newblock \emph{SIAM journal on numerical analysis}, 19\penalty0 (4):\penalty0
  742--760, 1982.

\bibitem[Babuška and Dorr(1981)]{babuska_hp_error_1981}
Ivo Babuška and Milo~R. Dorr.
\newblock Error estimates for the combined h and p versions of the finite
  element method.
\newblock \emph{Numerische Mathematik}, 37\penalty0 (2):\penalty0 257--277,
  1981.

\bibitem[Balay et~al.(1997)Balay, Gropp, McInnes, and
  Smith]{petscAutoGen_petsc-efficient}
Satish Balay, William~D. Gropp, Lois~Curfman McInnes, and Barry~F. Smith.
\newblock Efficient management of parallelism in object oriented numerical
  software libraries.
\newblock In E.~Arge, A.~M. Bruaset, and H.~P. Langtangen, editors,
  \emph{Modern Software Tools in Scientific Computing}, pages 163--202.
  Birkh{\"{a}}user Press, 1997.
\newblock \doi{10.1007/978-1-4612-1986-6_8}.

\bibitem[Balay et~al.(2018)Balay, Abhyankar, Adams, Brown, Brune, Buschelman,
  Dalcin, Eijkhout, Gropp, Kaushik, Knepley, May, McInnes, Mills, Munson, Rupp,
  Sanan, Smith, Zampini, Zhang, and Zhang]{petscAutoGen_petsc-user-ref}
Satish Balay, Shrirang Abhyankar, Mark~F. Adams, Jed Brown, Peter Brune, Kris
  Buschelman, Lisandro Dalcin, Victor Eijkhout, William~D. Gropp, Dinesh
  Kaushik, Matthew~G. Knepley, Dave~A. May, Lois~Curfman McInnes, Richard~Tran
  Mills, Todd Munson, Karl Rupp, Patrick Sanan, Barry~F. Smith, Stefano
  Zampini, Hong Zhang, and Hong Zhang.
\newblock {PETS}c users manual.
\newblock Technical Report ANL-95/11 - Revision 3.9, Argonne National
  Laboratory, 2018.
\newblock URL \url{http://www.mcs.anl.gov/petsc}.

\bibitem[Brezzi et~al.(1987)Brezzi, Douglas, Durán, and
  Fortin]{brezzi_mixed_1987}
Franco Brezzi, Jim Douglas, Ricardo Durán, and Michel Fortin.
\newblock Mixed finite elements for second order elliptic problems in three
  variables.
\newblock \emph{Numerische Mathematik}, 51\penalty0 (2):\penalty0 237--250,
  1987.

\bibitem[Chorin(1968)]{Chorin_1968}
Alexandre~Joel Chorin.
\newblock Numerical solution of the {Navier-Stokes} equations.
\newblock \emph{Mathematics of computation}, 22\penalty0 (104):\penalty0
  745–762, 1968.

\bibitem[Cockburn et~al.(2005)Cockburn, Kanschat, and
  Schötzau]{cockburn_locally_2005}
Bernardo Cockburn, Guido Kanschat, and Dominik Schötzau.
\newblock A locally conservative {LDG} method for the incompressible
  {Navier}-{Stokes} equations.
\newblock \emph{Mathematics of Computation}, 74\penalty0 (251):\penalty0
  1067--1096, 2005.

\bibitem[Dalcin et~al.(2011)Dalcin, Paz, Kler, and
  Cosimo]{petscAutoGen_Dalcin2011}
Lisandro~D. Dalcin, Rodrigo~R. Paz, Pablo~A. Kler, and Alejandro Cosimo.
\newblock Parallel distributed computing using {P}ython.
\newblock \emph{Advances in Water Resources}, 34\penalty0 (9):\penalty0
  1124--1139, 2011.
\newblock \doi{10.1016/j.advwatres.2011.04.013}.
\newblock New Computational Methods and Software Tools.

\bibitem[Davis(2004)]{petscAutoGen_davis2004algorithm}
Timothy~A Davis.
\newblock Algorithm 832: {UMFPACK} v4.3---an unsymmetric-pattern multifrontal
  method.
\newblock \emph{ACM Transactions on Mathematical Software (TOMS)}, 30\penalty0
  (2):\penalty0 196--199, 2004.
\newblock \doi{10.1145/992200.992206}.

\bibitem[Dean(1965)]{dean_1965}
Robert~G. Dean.
\newblock Stream function representation of nonlinear ocean waves.
\newblock \emph{Journal of Geophysical Research}, 70\penalty0 (18):\penalty0
  4561--4572, 1965.

\bibitem[Geuzaine and Remacle(2009)]{gmsh09}
Christophe Geuzaine and Jean-François Remacle.
\newblock Gmsh: A 3-d finite element mesh generator with built-in pre- and
  post-processing facilities.
\newblock \emph{International Journal for Numerical Methods in Engineering},
  79\penalty0 (11):\penalty0 1309--1331, 2009.

\bibitem[Green and Taylor(1937)]{green_mechanism_1937}
AE~Green and GI~Taylor.
\newblock Mechanism of the production of small eddies from larger ones.
\newblock In \emph{Proceedings of the royal society of {London}. {Series} {A},
  mathematical and physical sciences}, volume 158, pages 499--521, 1937.

\bibitem[Grue and Huseby(2002)]{grue_huseby_2002}
John Grue and Morten Huseby.
\newblock Higher-harmonic wave forces and ringing of vertical cylinders.
\newblock \emph{Applied Ocean Research}, 24\penalty0 (4):\penalty0 203--214,
  2002.

\bibitem[Harten(1983)]{harten_high_resolution_schemes_1983}
Ami Harten.
\newblock High resolution schemes for hyperbolic conservation laws.
\newblock \emph{Journal of Computational Physics}, 49\penalty0 (3):\penalty0
  357--393, 1983.

\bibitem[Hirt and Nichols(1981)]{hirt_volume_1981}
Cyril~W. Hirt and Billy~D. Nichols.
\newblock Volume of fluid ({VOF}) method for the dynamics of free boundaries.
\newblock \emph{Journal of Computational Physics}, 39\penalty0 (1):\penalty0
  201--225, 1981.

\bibitem[Huerta et~al.(2013)Huerta, Angeloski, Roca, and Peraire]{Huerta13}
Antonio Huerta, Aleksandar Angeloski, Xevi Roca, and Jaime Peraire.
\newblock Efficiency of high-order elements for continuous and discontinuous
  {Galerkin} methods.
\newblock \emph{International Journal for Numerical Methods in Engineering},
  96\penalty0 (9):\penalty0 529--560, 2013.

\bibitem[Issa and Violeau(2006)]{issa_ercoftac_2006}
Réza Issa and Damien Violeau.
\newblock {ERCOFTAC} test-case 2, 3d dambreaking, release 1.1, 2006.
\newblock URL \url{http://spheric-sph.org/tests/test-2}.

\bibitem[Kirby and Logg(2006)]{kirby_compiler_2006}
Robert~C. Kirby and Anders Logg.
\newblock A compiler for variational forms.
\newblock \emph{ACM Transactions on Mathematical Software}, 32\penalty0
  (3):\penalty0 417--444, September 2006.

\bibitem[Kirby et~al.(2012)Kirby, Sherwin, and Cockburn]{Kirby12}
Robert~M. Kirby, Spencer~J. Sherwin, and Bernardo Cockburn.
\newblock To {CG} or to {HDG}: A comparative study.
\newblock \emph{Journal of Scientific Computing}, 51\penalty0 (1):\penalty0
  183--212, 2012.

\bibitem[Kleefsman et~al.(2005)Kleefsman, Fekken, Veldman, Iwanowski, and
  Buchner]{kleefsman_vof_2005}
K.~M.~T. Kleefsman, G.~Fekken, A.~E.~P. Veldman, B.~Iwanowski, and B.~Buchner.
\newblock A volume-of-fluid based simulation method for wave impact problems.
\newblock \emph{Journal of Computational Physics}, 206\penalty0 (1):\penalty0
  363--393, 2005.

\bibitem[Krivodonova et~al.(2004)Krivodonova, Xin, Remacle, Chevaugeon, and
  Flaherty]{krivodonova_shock_2004}
L.~Krivodonova, J.~Xin, J.~F. Remacle, N.~Chevaugeon, and J.~E. Flaherty.
\newblock Shock detection and limiting with discontinuous {Galerkin} methods
  for hyperbolic conservation laws.
\newblock \emph{Applied Numerical Mathematics}, 48\penalty0 (3):\penalty0
  323--338, 2004.

\bibitem[Kubatko et~al.(2009)Kubatko, Bunya, Dawson, Westerink, and
  Mirabito]{Kubatko09}
Ethan~J. Kubatko, Shintaro Bunya, Clint Dawson, Joannes~J. Westerink, and Chris
  Mirabito.
\newblock A performance comparison of continuous and discontinuous finite
  element shallow water models.
\newblock \emph{Journal of Scientific Computing}, 40\penalty0 (1):\penalty0
  315--339, 2009.

\bibitem[Kuzmin(2010)]{kuzmin_vertex-based_2010}
Dmitri Kuzmin.
\newblock A vertex-based hierarchical slope limiter for $p$-adaptive
  discontinuous {Galerkin} methods.
\newblock \emph{Journal of Computational and Applied Mathematics}, 233\penalty0
  (12):\penalty0 3077--3085, April 2010.

\bibitem[Landet(2019{\natexlab{a}})]{landet_joss_2019}
Tormod Landet.
\newblock Ocellaris: a {DG} {FEM} solver for free-surface flows.
\newblock \emph{The Journal of Open Source Software}, 4\penalty0 (35):\penalty0
  1239, 2019{\natexlab{a}}.

\bibitem[Landet(2019{\natexlab{b}})]{landet_ocellaris_user_guide_2019}
Tormod Landet.
\newblock Ocellaris web page and user manual, 2019{\natexlab{b}}.
\newblock \href{https://www.ocellaris.org/}{www.ocellaris.org}.

\bibitem[Landet(2019{\natexlab{c}})]{landet_p3_zenodo_2019}
Tormod Landet.
\newblock Ocellaris {DG}-{FEM} software and input files to reproduce results,
  01 2019{\natexlab{c}}.
\newblock \href{http://doi.org/10.5281/zenodo.2587038}{Zenodo:
  10.5281/zenodo.2587038}.

\bibitem[Landet and Mortensen(2019)]{landet_pcorr_2019}
Tormod Landet and Mikael Mortensen.
\newblock On exactly incompressible {DG} {FEM} pressure splitting schemes for
  the {Navier}-{Stokes} equation.
\newblock \emph{{arXiv}:\href{http://arxiv.org/abs/1903.11943}{1903.11943
  [physics]}}, 2019.
\newblock URL \url{http://arxiv.org/abs/1903.11943}.

\bibitem[Landet et~al.(2018)Landet, Mardal, and
  Mortensen]{landet_slopelim_2019}
Tormod Landet, Kent-Andre Mardal, and Mikael Mortensen.
\newblock Slope limiting the velocity field in a discontinuous {Galerkin}
  divergence free two-phase flow solver.
\newblock \emph{{arXiv}:\href{http://arxiv.org/abs/1803.06976}{1803.06976
  [physics]}}, 2018.
\newblock URL \url{http://arxiv.org/abs/1803.06976}.

\bibitem[LLNL()]{petscAutoGen_hypre-web-page}
LLNL.
\newblock {\sl hypre}: High performance preconditioners.
\newblock \href{http://www.llnl.gov/CASC/hypre/}{www.llnl.gov/CASC/hypre/}.

\bibitem[Logg et~al.(2012)Logg, Mardal, and Wells]{logg_automated_2012}
Anders Logg, Kent-Andre Mardal, and Garth Wells.
\newblock \emph{Automated Solution of Differential Equations by the Finite
  Element Method: The {FEniCS} Book}.
\newblock Springer Science \& Business Media, February 2012.

\bibitem[Muzaferija et~al.(1998)Muzaferija, Peric, Sames, and
  Schellin]{muzaferija_hric_1998}
Samir Muzaferija, Milovan Peric, Pierre~C. Sames, and Thomas~E. Schellin.
\newblock A two-fluid {Navier}-{Stokes} solver to simulate water entry.
\newblock In \emph{Proceedings from the 22nd symposium on naval hydrodynamics},
  pages 277--289, Washington, DC, 1998.

\bibitem[Paulsen et~al.(2014)Paulsen, Bredmose, Bingham, and
  Jacobsen]{paulsen_forcing_2014}
Bo~T. Paulsen, H.~Bredmose, H.~B. Bingham, and N.~G. Jacobsen.
\newblock Forcing of a bottom-mounted circular cylinder by steep regular water
  waves at finite depth.
\newblock \emph{Journal of Fluid Mechanics}, 755:\penalty0 1--34, 2014.

\bibitem[Perić and Abdel-Maksoud(2016)]{peric_reliable_2016}
R.~Perić and M.~Abdel-Maksoud.
\newblock Reliable damping of free-surface waves in numerical simulations.
\newblock \emph{Ship Technology Research}, 63\penalty0 (1):\penalty0 1--13,
  January 2016.

\bibitem[Perić and Abdel-Maksoud(2018)]{peric_analytical_2018}
Robinson Perić and Moustafa Abdel-Maksoud.
\newblock Analytical prediction of reflection coefficients for wave absorbing
  layers in flow simulations of regular free-surface waves.
\newblock \emph{Ocean Engineering}, 147:\penalty0 132--147, 2018.

\bibitem[Popinet(2003)]{popinet_gerris_2003}
Stéphane Popinet.
\newblock Gerris: a tree-based adaptive solver for the incompressible {Euler}
  equations in complex geometries.
\newblock \emph{Journal of Computational Physics}, 190\penalty0 (2):\penalty0
  572--600, 2003.

\bibitem[Rienecker and Fenton(1981)]{rienecker_fenton_1981}
M.~M. Rienecker and J.~D. Fenton.
\newblock A {Fourier} approximation method for steady water waves.
\newblock \emph{Journal of Fluid Mechanics}, 104:\penalty0 119--137, 1981.

\bibitem[Temam(1969)]{Temam_1969}
Roger Temam.
\newblock Sur l’approximation de la solution des équations de
  {Navier-Stokes} par la méthode des pas fractionnaires (ii).
\newblock \emph{Archive for rational mechanics and analysis}, 33\penalty0
  (5):\penalty0 377–385, 1969.

\bibitem[Ubbink(1997)]{ubbink_1997}
Onno Ubbink.
\newblock Numerical prediction of two fluid systems with sharp interfaces,
  1997.

\bibitem[Weller et~al.(1998)Weller, Tabor, Jasak, and Fureby]{weller_FOAM_1998}
H.~G. Weller, G.~Tabor, H.~Jasak, and C.~Fureby.
\newblock A tensorial approach to computational continuum mechanics using
  object-oriented techniques.
\newblock \emph{Computers in Physics}, 12\penalty0 (6):\penalty0 620--631,
  1998.

\bibitem[Ølgaard et~al.(2008)Ølgaard, Logg, and
  Wells]{olgaard_automated_2008}
K.~Ølgaard, A.~Logg, and G.~Wells.
\newblock Automated code generation for discontinuous {Galerkin} methods.
\newblock \emph{SIAM Journal on Scientific Computing}, 31\penalty0
  (2):\penalty0 849--864, November 2008.

\end{thebibliography}

\end{document}